\documentclass[12pt,aps,epsf,twocolumn,numberedappendix,appendixfloats]{emulateapj}

\usepackage[breaklinks,colorlinks,urlcolor=blue,citecolor=blue,linkcolor=blue]{hyperref}
\usepackage{savesym}
\savesymbol{tablenum}
\usepackage{siunitx}
\restoresymbol{SIX}{tablenum}
\usepackage{mathtools}
\usepackage{physics}
\usepackage{graphicx}
\usepackage[FIGTOPCAP]{subfigure}
\usepackage{rotating}
\usepackage{epstopdf}
\usepackage{dcolumn}
\usepackage{bm}
\usepackage{enumitem} 
\usepackage{makecell}

\def\msun{{M_\odot}}

\def\beq{\begin{equation}}
\def\eeq{\end{equation}}
\newcommand{\eq}[1]{\begin{align}#1\end{align}}
\def\barr{\begin{eqnarray}}
\def\earr{\end{eqnarray}}
\def\bal{\begin{align}}
\def\eal{\end{align}}

\def\bsub{\begin{subequations}}
\def\esub{\end{subequations}}

\def\SMBH{SMBH}
\def\mSMBH{M_\mathrm\SMBH}

\def\mSMBHmin{{M_\mathrm{SMBH,min}}}
\def\mSMBHmax{{M_\mathrm{SMBH,max}}}

\def\SMBH{SMBH}

\def\mbh{m_\mathrm{BH}}

\def\rdf{r_\mathrm{DF}}
\def\rgw{r_\mathrm{GW}}

\def\tdf{t_\mathrm{DF}}
\def\tdfo{t_\mathrm{DF,0}}
\def\ninit{N_\mathrm{init}}
\def\ntot{N_\mathrm{tot}}

\def\fbhi{f_\mathrm{BH,init}(m)}
\def\rhost{\rho_\ast}
\def\mmin{{m_\mathrm{min}}}
\def\mmax{{m_\mathrm{max}}}
\def\rmin{{r_\mathrm{min}}}

\def\tgw{{t_\mathrm{GW}}}

\def\nbh{{N_\mathrm{BH}}}
\def\mtot{{m_\mathrm{tot}}}
\def\mcr{{m_\mathrm{cr}}}
\def\cinf{{C_\mathrm{inf}}}
\def\ngal{{n_\mathrm{gal}}}
\def\ngw{{N_\mathrm{GW}}}
\def\XI{\zeta}

\begin{document}

\title{The rate of stellar mass black hole scattering in galactic nuclei}

\author{Alexander Rasskazov and Bence Kocsis}
\affil{E\"otv\"os University, Institute of Physics, P\'azm\'any P. s. 1/A, Budapest, Hungary 1117}

\begin{abstract}
We consider a black hole (BH) density cusp in a nuclear star cluster (NSC) hosting a supermassive back hole (SMBH) at its center. Assuming the stars and BHs inside the SMBH sphere of influence are mass-segregated, we calculate the number of BHs that sink into this region under the influence of dynamical friction. We find that the total number of BHs increases significantly in this region due to this process for lower mass SMBHs by up to a factor of 5, but there is no increase in the vicinity of the highest mass SMBHs. Due to the high BH number density in the NSC, BH-BH binaries form during close approaches due to GW emission. We update the previous estimate of O'Leary et al. for the rate of such GW capture events by estimating the $\langle n^2\rangle/\langle n\rangle^2$ parameter where $n$ is the number density. We find a BH merger rate for this channel to be in the range $\sim0.002-0.04\,\si{Gpc^{-3}yr^{-1}}$. The total merger rate is dominated by the smallest galaxies hosting SMBHs,  
and the number of heaviest BHs in the NSC. It is also exponentially sensitive to the radial number density profile exponent, reaching $>\SI{100}{Gpc^{-3}yr^{-1}}$ when the BH mass function is $m^{-2.3}$ or shallower and the heaviest BH radial number density is close to $r^{-3}$. 
Even if the rate is much lower than the range constrained by the current LIGO detections, the GW captures around SMBHs can be distinguished by their high eccentricity in the LIGO band. 
\end{abstract}

\section{Introduction}

Ten stellar black hole - black hole (BH-BH) detections of binary mergers have been announced to date by Advanced LIGO and Virgo, which implies a merger rate density of $24-112\,\si{yr^{-1}Gpc^{-3}}$ in the Universe for a power law BH mass function prior \citep{ligo2018,ligo2018a}.
Several astrophysical channels have been proposed to explain these rates including isolated binary evolution in the galactic field \citep{Belczynski2016} and dynamically formed binaries in globular clusters \citep{Rodriguez2016,Fragione2018}. All these events are consistent with being approximately circular, which is expected if the binaries form with a sufficiently large periapsis, since gravitational wave (GW) emission circularizes the orbit as it shrinks \citep{Peters1964}. 

However, a few channels are predicted to produce eccentric BH binaries that retain significantly nonzero eccentricity in the Advanced LIGO frequency range ($\gtrsim\SI{10}{Hz}$ for design sensitivity). First, the Kozai-Lidov effect can enhance the BH binary eccentricity in hierarchical triples. The tertiary component can be a star or another stellar mass BH. Such triples can form in the galactic field \citep{Antonini2017,Silsbee2017} or in globular clusters as a result of a binary-binary interaction \citep{Antonini2016}. Alternatively, the tertiary component may be a supermassive BH (SMBH) in a galactic center \citep{Hoang2018,Antonini2012,Hamers2018}
Furthermore, ``resonant'' binary-single scattering interactions may lead to highly eccentric binaries in globular clusters where the tertiary increases the eccentricity of the inner binary during close pericenter passages \citep{Samsing2018,Rodriguez2018}. Eccentric BH binary GW sources can also be created from non-hierarchical triples \citep{Arca-Sedda2018}.

The focus of this paper is on another way to produce highly eccentric BH--BH binaries, the so-called ``GW captures'' 
in which two single BHs undergo a close encounter and lose a sufficient amount of energy due to GW emission to become bound \citep{OLeary2009}. For a sufficiently low impact parameter, the newly-formed binary has a sufficiently small semimajor axis and high eccentricity to merge quickly before it is disrupted by an interaction with another star or BH. These events are most frequent in dense stellar clusters, e.g. galactic nuclei and globular clusters. However, the low relative velocity of BHs in globular clusters implies that most GW captures will form binaries on a wide orbit, and the binary eccentricity will typically become low due to GW emission when reaching the LIGO band \citep[][Figure 6]{OLeary2009}. In contrast, BHs in galactic nuclei sink to the inner regions close to the supermassive black hole (SMBH) due to dynamical friction, and form a mass segregated steep density cusp, where the velocity dispersion is much higher \citep{BW77,FAK06,HA06,OLeary2009}. In these environments, GW capture binaries typically form in the LIGO band with high eccentricities \citep{Gondan2017}. As was shown in \citet{Gondan2018}, the aLIGO-adVirgo-KAGRA detector network will be able to measure the merging BH binary's eccentricity with high accuracy and therefore potentially distinguish the GW capture events from other astrophysical formation channels.

The purpose of this paper is to refine the rate estimate of GW capture events in galactic nuclei, highlight the main sources of uncertainties, and to calculate the distribution of their total masses and mass ratios. Previously, \citet{OLeary2009} has estimated the event rates by solving the Fokker-Planck equations for isotropic multimass BH distributions. The results were quite different for models with a limited mass range of stellar BHs and models with a BH mass function that extends to higher masses (e.g. $m_{\max}=15\msun$ vs. $45\msun$). Importantly, the rates were found to be proportional to a parameter $\xi$ defined as the mean squared number density over the square of the mean number density of galactic nuclei. The value of this parameter was not estimated, its fiducial value was assumed to be $\xi=30$. \citet{Kocsis2012} extended \citet{OLeary2009} using post-Newtonian simulations, and showed with simple analytical estimates that rare galaxies with a high $\xi$ and BH mass fractions may dominate the rates, but did not determine these quantities. \citet{Tsang2013} estimated $\xi$, but did not calculate BH-BH merger rates, but rather focused on NS-NS binaries which do not form highly mass segregated density cusps. \citet{Gondan2017} derived the mass and eccentricity distribution of GW capture binaries, but also did not estimate their total merger rate. 

In this paper we fill in the missing pieces in the puzzle to determine the GW capture rate using simple analytical estimates and examine the dependence on various model parameters. We estimate $\xi$ based on the observed scatter of the $M-\sigma$ relation. Further, we calculate the BH number density taking into account 
the dynamical friction bringing BHs (predominantly heavier ones) into the galactic center \citep{Miralda2000}. 
Given the number density, the event rate does not depend on any additional parameters \citep{OLeary2009,Gondan2017}. We consider various assumptions for 
the initial BH mass function and,{ based on the heaviest BH detected by LIGO \citep{ligo2018}, assume it extends up to $50\msun$.}
We also briefly consider the effect of the steepness of the BH density cusp \citep{Keshet2009}.

The paper is organized as follows. In Section~\ref{section:number-of-bhs}, we calculate the number of BHs around the SMBH using the results of previous papers about their mass-segregated density profile. Then in Section~\ref{section:mergerrate} we utilize this result to calculate the rate of GW captures in a galactic nucleus. Finally, in Section~\ref{section:total-merger-rate} we integrate over all galaxies and calculate the event rate per unit volume. In Section~\ref{section:conclusions} we summarize our conclusions and briefly discuss other eccentric BH merger mechanisms. {Several details about the calculations are given in the appendix.}

\section{Number of BHs in a galactic center}
\label{section:number-of-bhs}

In this section we calculate the increase in the number of stellar-mass BHs within the SMBH radius of influence $r_0$
due to the sinking of BHs from larger radii caused by dynamical friction. 

\subsection{Initial conditions}

First, we assume that all stars and BHs formed in the galactic center early 
in the galactic lifetime ($T=\SI{12}{Gyr}$ ago) and that every star heavier than a certain mass $\mcr$ produced a BH remnant. We will also consider the case of continuous star formation in the next subsection. The initial stellar mass function is taken from \cite{KroupaIMF}:
\beq\label{eq:KroupaIMF} 
f_\mathrm{IMF} (m) = C \begin{dcases} 
25\left(\frac{m}{\msun}\right)^{-0.3}\qc m<0.08\msun\,,\\
2\left(\frac{m}{\msun}\right)^{-1.3}\qc 0.08\msun<m<0.5\msun\,,\\
\left(\frac{m}{\msun}\right)^{-2.3}\qc m>0.5\msun,
\end{dcases}
\eeq
where $C$ is a normalization parameter.
The total stellar mass is then
\barr\label{eq:mstellar}
M_\ast = \int_0^\infty m f_\mathrm{IMF} (m) \dd{m} = 5.58M_\odot^2 C
\earr
and the total (initial) number of BHs is 
\begin{align}
\ninit &= \int_{\mcr}^\infty f_\mathrm{IMF} (m) \dd{m} = 
k M_\ast.
\label{eq:ninit}
\end{align}
where
\beq\label{eq:k}
k = \num{2.86e-3} M_\odot^{-1} \qty(\frac{\mcr}{20\msun})^{-1.3}
\eeq
Eq. 
(\ref{eq:ninit}) allows us to calculate the initial number of BHs in any region where we know the total stellar mass; for example, inside the influence radius $r_0$ defined where $M_\ast=2\mSMBH$
\barr\label{eq:ninitr0}
\ninit (r_0) &=& 2k\mSMBH.
\earr
Thus, the initial BH mass fraction is 
\beq
\kappa_\mathrm{init} = \frac{\ninit \langle m_{\rm BH}\rangle}{M_\ast} = k \langle m_{\rm BH}\rangle
\eeq
where $\langle m_{\rm BH}\rangle$ is the average BH mass. Given a power-law BH mass distribution 
\begin{equation}\label{eq:fBH}
 \fbhi\propto m_\mathrm{BH}^{-\beta}\,,~~\mmin<m_\mathrm{BH}<\mmax   
\end{equation}
the average BH mass is
\beq\label{eq:<mbh>}
\langle m_{\rm BH}\rangle = 
\frac{\int_\mmin^\mmax m^{1-\beta}\dd{m}}{\int_\mmin^\mmax m^{-\beta}\dd{m}} = \frac{\beta-1}{\beta-2} \cdot\frac{\mmin^{2-\beta}-\mmax^{2-\beta}}{\mmin^{1-\beta}-\mmax^{1-\beta}}.
\eeq
For example, $\mmin=5\msun$ and $\mmax=40\msun$ give $\kappa_\mathrm{init} \approx 0.03$ for $2<\beta<3$. 

The mass distribution of BHs born inside $r_0$ 
is
\begin{align}
\dv{\ninit}{m} &= \fbhi k M_\ast(r_0) 
\nonumber\\&= \fbhi k \int_0^{r_0} \rho_\ast(r)\,4\pi r^2\dd{r},
\end{align}
where $\rho_\ast(r)$ is stellar density:
\beq\label{eq:rhoinside} 
\rhost(r) = \begin{dcases} 
\rho_0 \left(\frac{r}{r_0}\right)^{-\gamma_1}\qc r\leq r_0\,,\\
\rho_0 \left(\frac{r}{r_0}\right)^{-\gamma_2}\qc r>r_0\,,
\end{dcases}
\eeq
where
\begin{equation}
\rho_{0} = \frac{(3-\gamma_1)}{4\pi}\frac{2\mSMBH}{ r_0^3}.\label{eq:rho0}
\end{equation}
This gives
\beq
\dv{\ninit}{m} = \frac{4\pi}{3-\gamma_1} k\rho_0r_0^3 \fbhi  
\eeq
As we consider the Milky Way (MW) NSC to be relaxed \citep{BW77}, we assume $\gamma_1=1.5$, which is consistent with observed deep star counts\footnote{However,  diffuse light measurements of \citet{Schodel2018} give a lower value $\gamma_1=1.13\pm0.08$. 
The difference in the BH number between $\gamma_1=1.5$ and $\gamma_1=1.1$ is only $\sim15\%$.} \citep{GallegoCano2018}. For the density profile outside $r_0$, we assume $\gamma_2=3.2$ which is consistent with both star counts and diffuse light measurements \citep{GallegoCano2018,Schodel2018}. 

\subsection{The effect of dynamical friction on the BH number density}

The black hole mass function in the NSC is affected by dynamical friction, which delivers BHs into this region. The total number of BHs with a given mass within $r_0$ 
at present is defined by the maximum radius $\rdf$ from where a BH sinks to within $r_0$ 
in a Hubble time:
\barr\label{eq:ntot}
\dv{\ntot}{m} &=& \fbhi \int_0^{\rdf(m)} n(r)\,4\pi r^2\dd{r}.
\earr
where $n(r)$ is the BH number density.
Here $\rdf(m)$ can be defined as the initial orbital radius of a BH given the final radius $r_0$ 
and BH mass $m$. The evolution of a BH orbital radius can be approximated as
\citep{BinneyTremaine}
\beq
\label{eq:drdt}
\dv{r}{t} \equiv -\frac{r}{\tdf} = -r\cdot \ln{\Lambda} \frac{4\pi G^2\rhost m}{\upsilon^3} \int_0^\upsilon 4\pi u^2F(u)\dd{u},
\eeq
where 
\beq
\ln\Lambda \approx \ln(M_{\bullet}/m) \approx 13, \label{eq:lnLambda}
\eeq
$\mSMBH$ is the central SMBH mass, $\upsilon$ is the BH velocity and $F(u)$ is the velocity distribution of ambient stars. 
For a Maxwellian velocity distribution 
the value of the integral is 0.54 for $v^2 = \langle u^2\rangle$.
Assuming a circular BH orbit,
\begin{align}
v = \sqrt\frac{GM(r)}{r} 
\end{align}
where $M(r)$ is the total mass inside of radius $r$:
\begin{align}
    M(r) &= 3\mSMBH + \int_{r_0}^{r} \rho_0 \qty(\frac{r}{r_0})^{-\gamma_2} 4\pi r^2 \dd{r} \nonumber\\
    &= \mSMBH{}\qty[3+2\frac{3-\gamma_1}{\gamma_2-3}\qty(1-\qty(\frac{r}{r_0})^{3-\gamma_2})].
\end{align}
As a result, the dependence of dynamical friction (DF) timescale on radius is the following:
\bsub
\begin{align}
\tdf &= \tdfo \, x^{\gamma_2-3/2} \qty[1+\frac{2}{3}\frac{3-\gamma_1}{\gamma_2-3}(1-x^{3-\gamma_2})]^{3/2},\\
x &\equiv \frac{r}{r_0},\\
\tdfo &\equiv \frac{(3r_0)^{3/2}M_\mathrm{SMBH}^{1/2}}{1.08(3-\gamma_1)\ln\Lambda \, G^{1/2}m}\nonumber\\
&= \SI{3.8}{Gyr}\, \frac{1.5}{3-\gamma_1} \qty(\frac{m}{10\,\msun})^{-1} \qty(\frac{r_0}{\SI{3}{pc}})^{3/2} \nonumber\\ 
&\times \qty(\frac{\mSMBH}{\num{4e6}\msun})^{1/2}. \label{eq:tdf0}
\end{align}
\esub

The equation of motion (\ref{eq:drdt}) then becomes
\bsub\label{eq:dxdtau}
\begin{align}
    \dv{x}{\tau} &= -x \qty(\frac{\tdf}{\tdfo})^{-1} \nonumber\\
    &= -x^{5/2-\gamma_2} \qty[1+\frac{2}{3}(3-\gamma_1)\frac{1-x^{3-\gamma_2}}{\gamma_2-3}]^{-3/2},\\
    \tau &\equiv \frac{t}{\tdfo}.
\end{align}
\esub

{The radius $\rdf(m)$ from which objects of mass $m$ sink to within the radius of influence $r_0$ within time $T$ satisfies}
\bsub\label{eq:boundary_conditions}
\begin{align}
    x\qty(\tau=\frac{T}{\tdfo}) &= 1,\\
    x(\tau=0) &= \frac{\rdf}{r_0}.
\end{align}
\esub

The numerical solution of Eq. \eqref{eq:dxdtau} with boundary conditions \eqref{eq:boundary_conditions} for $\gamma_1=1.5$, $\gamma_2=3.2$ can be approximated (with $3\%$ accuracy for $5\msun<m<40\msun$) as
\begin{align}\label{eq:rdf-outside}
\rdf &= r_0 \qty(1+k_1\qty(\frac{T}{\tdfo})^{k_2})^{k_3},
\end{align}
where $k_1=1.016$, $k_2=0.740$, $k_3=0.654$.

\begin{figure}
	\centering
	\subfigure{\includegraphics[width=0.49\textwidth]{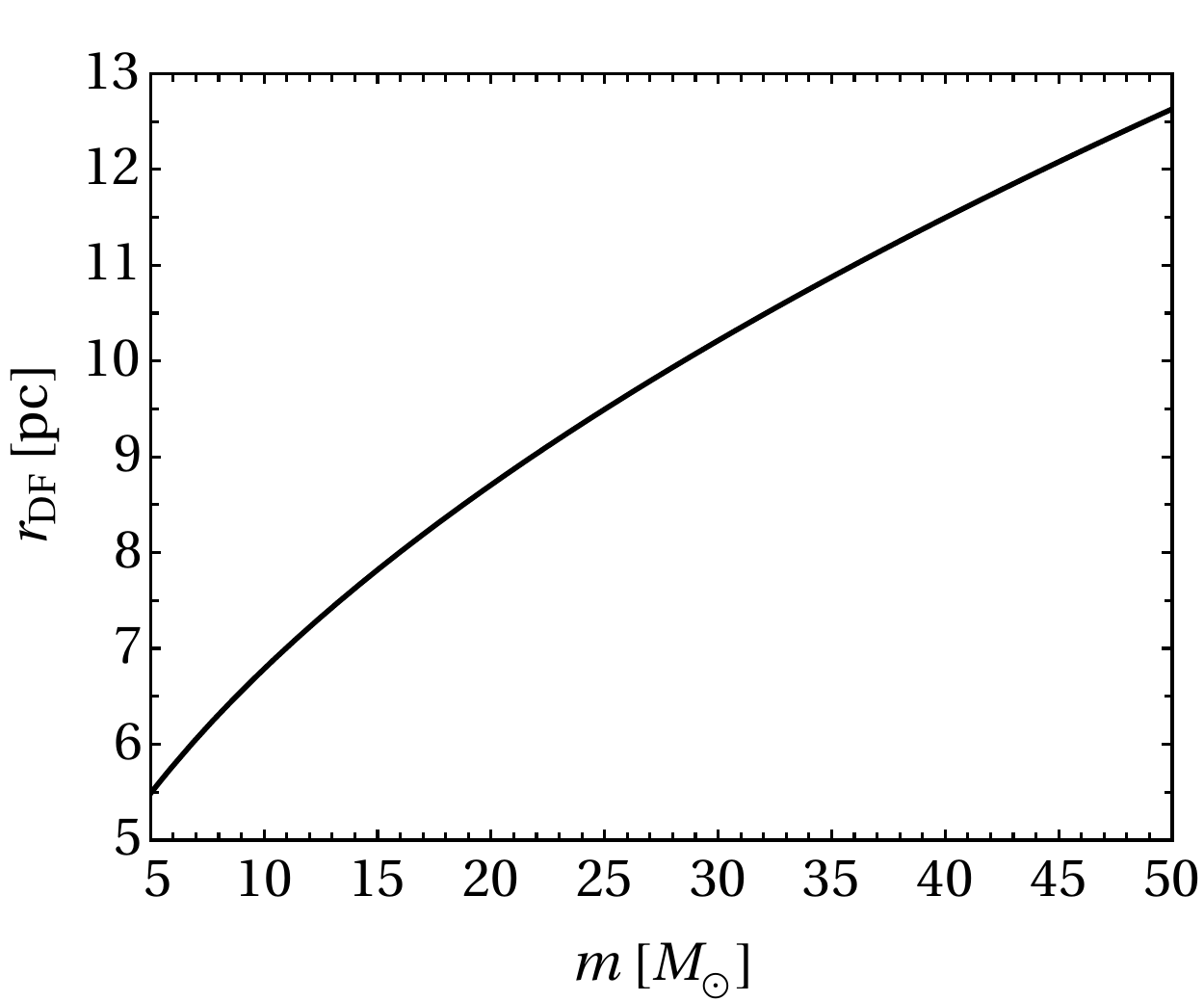}}
\caption{
The radius from where the BHs with mass $m$ would sink inside $r_0$ for a MW-like galaxy.  
}
\label{fig:rdf}
\end{figure}

{Note that $\rdf$ depends on $m$ through $\tdfo$, as shown on Fig.~\ref{fig:rdf}. E.g in a MW-like galaxy, for $m=40\,\msun$ $\rdf\approx\SI{11}{pc}$ which is about the distance where the NSC starts dominating over the galactic background in the MW \citep[if we take the MW stellar density from e.g.][]{Gnedin2014}. }

After we substitute 
Eq. (\ref{eq:rdf-outside}) into Eq.~(\ref{eq:ntot}), we find that the BH mass function within distance $r_0$ after a Hubble time is
\begin{align}
\dv{\ntot}{m} &= 
4\pi k\rho_0 r_0^3\fbhi \nonumber\\
&\times \left\{\frac{1}{3-\gamma_1} + 
\frac{1}{\gamma_2-3}\qty[\qty(1-\frac{\rdf(m)}{r_0})^{3-\gamma_2}]\right\}.
\end{align}
As $\fbhi$ and $k$ (Eq.~\ref{eq:k}) are highly uncertain, it is useful to calculate the relative increase in the number of BHs due to DF:
\eq{\label{eq:rdf}
\XI &\equiv \frac{\dv*{\ntot}{m}}{\dv*{\ninit}{m}} = 
1 + \frac{3-\gamma_1}{\gamma_2-3}\qty(1-\qty(\frac{\rdf(m)}{r_0})^{3-\gamma_2}) \nonumber\\
&= 1 + \frac{3-\gamma_1}{\gamma_2-3}
\qty[1-\qty{1+k_1\qty(\frac{T}{\tdfo})^{k_2}}^{(3-\gamma_2)k_3}]
}

\begin{figure}
	\centering
	\subfigure{\includegraphics[width=0.49\textwidth]{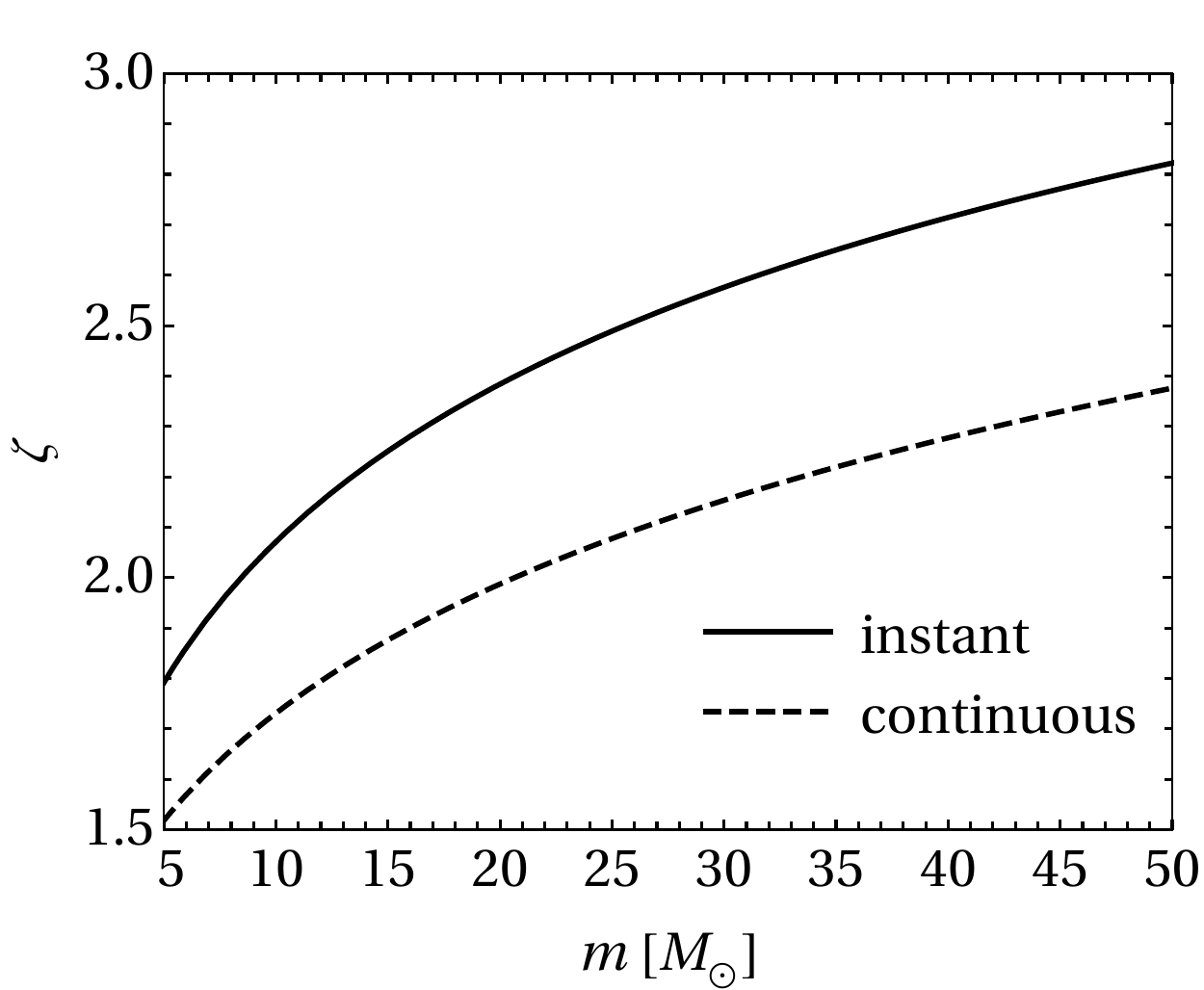}}
\caption{
The relative increase in the number of BHs in the center of a Milky Way-like galaxy due to dynamical friction depending on the BH mass $m$. All BHs are assumed to form 12 Gyr ago (solid) or with a constant rate during the last 12 Gyr (dashed).
The influence radius is $r_0=\SI{2.45}{pc}$ (Eq. \ref{eq:r_0}) and the inner and outer stellar density slopes are $\gamma_1=1.5$, $\gamma_2=3.2$, respectively. 
}
\label{fig:xi}
\end{figure}


So far we have assumed that all of the BHs in the Galactic center were born $T=\SI{12}{Gyr}$ ago. Additionally, we also consider the possibility in which the BHs are produced continuously in time with a constant rate. 
The true BH formation history may be expected to lie between these two extreme cases, and the corresponding estimates for the number of BHs or their merger rate may represent upper and lower limits, respectively. The corresponding expressions for $\XI$ are derived in Appendix~\ref{appendix:xi-continuous}. 

Figure~\ref{fig:xi} shows that the value of $\XI$ ranges between 1.5 and 3 in MW-like galaxies depending on the BH mass and the assumed BH formation history. However, as shown in the next section, $\XI$ is higher or lower in more or less massive galaxies, respectively.

\section{Merger rate per galaxy}
\label{section:mergerrate}

The rate of GW captures in a single NSC, $\Gamma$, may be calculated by adding up the contribution of different radial shells around the SMBH {within its radius of influence}, accounting for the local flux of objects and the cross section to form binaries by GW emission for given BH masses. To do that, we adopt the formula from \citet[][Eq. 125]{Gondan2017}:
\begin{align}\label{eq:gamma0mamb}
 \frac{\partial^2\Gamma_0}{\partial m_A \partial m_B}   \approx& \frac{G^{17/14}}{c^{10/7}}
 \frac{N_{\rm BH}^2 c_\eta^{2/7}}{\mSMBH^{11/14}\,r_0 ^{31/14}}
 \nonumber \\ & \times
 \frac{ 9 m_\mathrm{max}^2 - 6 p_0 m_\mathrm{max} m_\mathrm{tot}
 + 4 p_0^2 \mu m_\mathrm{tot} }{16\, m_\mathrm{max}^2 }
 \nonumber \\ & \times 
 \frac{ m_\mathrm{tot}^{2-\beta} \mu^{-\beta} (1-\beta)^2 }{  \qty(m_\mathrm{max}
 ^{1-\beta} - m_\mathrm{min}^{1-\beta} )^2} 
  \nonumber \\ & \times 
  \frac{ 1 - (r_\mathrm{min}/r_0)^{ 11/14 - p_0 m_\mathrm{tot} / m_\mathrm{max} 
  } }{\frac{11}{14} - p_0 \frac{ m_\mathrm{tot} }{ m_\mathrm{max} } }.
\end{align}

Here $m_{A,B}$ are the BH masses, $\rmin$ and $r_0$ are the minimum and maximum radii of the BH density distribution, $N_{\rm BH}$ is the total number of BHs within $\rmin<r<r_0$, $\beta$ is the BH mass function power-law (as in Eq.~\ref{eq:fBH}) and $m_{\rm min,\,max}$ are the minimum and maximum BH masses (in this paper we assume $\mmin=5\msun$, $\mmax=40\msun$). Also, 
\begin{equation}
  \mtot \equiv m_A+m_B,\quad
  \mu \equiv \frac{m_Am_B}{m_A+m_B},\quad
  \eta \equiv \frac{\mu}{\mtot}
  \end{equation}
are the total mass, reduced mass, and symmetric mass ratio, respectively, and $c_\eta \equiv (340\pi/3)\eta$.

{
For illustrative purposes, we also derive the formulae for the rates of mergers between the smallest and the heaviest BHs as
\bsub
\eq{
\Gamma_{\min} &= \left.\frac{\partial^2 \Gamma_0}{\partial\ln m_A\, \partial \ln m_B}\right|_{m_A=m_B=m_{\min}} \nonumber\\
&= \frac{\partial^2\Gamma_0}{\partial m_A \partial m_B} m_{\min}^2 \nonumber\\
&\approx C \cdot \frac{63}{22} \qty[\frac{\beta-1}{1-\qty(\frac{\mmin}{\mmax})^{\beta-1}}]^2 m_{\min}^2,\\
\Gamma_{\max} &= \left.\frac{\partial^2 \Gamma_0}{\partial\ln m_A\, \partial \ln m_B}\right|_{m_A=m_B=m_{\max}} \nonumber\\
&\approx C \qty[\frac{\beta-1}{1-\qty(\frac{\mmin}{\mmax})^{1-\beta}}]^2 \frac{14}{3}\qty(\frac{\rmin}{r_0})^{-3/14} m_{\max}^2,\\
C &\equiv \frac{G^{17/14}N_{\rm BH}^2 c_\eta^{2/7}}{c^{10/7}\mSMBH^{11/14}\,r_0 ^{31/14}}.
}
\esub
Here we assume $p_0=0.5$ and $\mmin\ll\mmax$. As we can see from those expressions, the majority of mergers happen mergers between the smallest BHs and the heaviest BHs when $\beta\gtrsim2$ and $\beta\lesssim2$, respectively (which is later shown more rigorously on Fig.~\ref{fig:mergerrate}):
\eq{
\frac{\Gamma_{\min}}{\Gamma_{\max}} \approx \frac{27}{44} \qty(\frac{\mmin}{\mmax})^{4-2\beta} \qty(\frac{\rmin}{r_0})^{3/14} .
}
}

Eq. (\ref{eq:gamma0mamb}) assumes a steady state mass-segregated radial 3D number density profile derived by \citet{OLeary2009} using the Fokker-Planck equation:
\beq\label{eq:n(m,r)}
n(m,r)\propto r^{-\frac{3}{2}-p_0\frac{m}{\mmax}},
\eeq
where $p_0\approx 0.5$ \citep{OLeary2009}, i.e. the radial power law index varies from  $-1.5$ for the lightest BHs to $-2$ for the heaviest ones. {However, we also examine different assumptions in Section~\ref{section:total-merger-rate}.}

\begin{table*}
\caption{Different assumptions for the minimum radius of BH distribution and the inferred BH-BH merger rates}
\centering
\begin{tabular}{l c c c}
\hline
Reference & \cite{OLeary2009} & \cite{Kocsis2012} & \cite{Gondan2017} \\ [0.5ex] 
\hline
$\rmin$ 
& 
\begin{tabular}{@{}c@{}}
$\tgw(\rmin) = t_H$ \\[1ex]
$\rmin \propto \mSMBH^{1/2}$
\end{tabular}
& 
\begin{tabular}{@{}c@{}}
$\nbh(\rmin) = 1$ \\[1ex]
$\rmin \propto \mSMBH^{-1/2}$
\end{tabular}
& 
\begin{tabular}{@{}c@{}}
$\tgw(\rmin) = t_\mathrm{rel}(\rmin)$ \\[1ex]
$\rmin \propto \mSMBH^{13/16}$
\end{tabular}
\\ [3ex]
\hline 
$\Gamma$ 
& 
$\propto \mSMBH^{3/28}$
& 
$\propto \mSMBH^{9/28}$
& 
\begin{tabular}{@{}c@{}}
$\propto \mSMBH^{3/28},\quad m_\mathrm{tot} < \frac{11}{7} m_\mathrm{max}$ \\[1ex]
$\propto \mSMBH^{9/224},\quad m_\mathrm{tot} > \frac{11}{7} m_\mathrm{max}$
\end{tabular}
\\ [3ex]
\hline 
\\ 
\end{tabular}
\label{table:comparison}
\end{table*}
  
As shown in \cite{Gondan2017}, for any $\mSMBH\lesssim10^7\msun$ the relaxation time inside $r_0$ is shorter than the Hubble time. Given that most of the merger events come from the low-mass galaxies \citep{OLeary2009}, this justifies {our assumption that the SBH sphere of influence is fully relaxed.} And since the BH density outside $r_0$ falls down quickly ($\propto r^{-3.2}$), we assume we can ignore the contribution of those BHs to the total merger rate.

Following \cite{Gondan2017}, we define $\rmin$ as the radius where the GW inspiral time becomes shorter than the relaxation time: 
\bsub
\barr
t_\mathrm{rel} &=& 0.34\frac{\sigma^3(r_\mathrm{min})}{G^2n(\rgw)\langle m^2\rangle\ln\Lambda}
\nonumber\\
&=& t_\mathrm{GW} = \frac{5c^5r_{\rm min}^4}{64G^3\mbh\mSMBH},\\
\rmin &=& \qty[
\frac{82}{\XI\ln\Lambda} r_0^{1.12} \qty(\frac{G\mSMBH}{c^2})^{2.5} \frac{\mbh}{\msun} \qty(\frac{\mcr}{20\msun})^{1.3}
]^{1/3.62}
\nonumber\\ 
&=& \SI{6.9e-5}{pc}\,\qty(\frac{\mSMBH}{\num{4e6}\msun})^{0.69} \qty(\frac{r_0}{\SI{3}{pc}})^{0.31} 
\nonumber\\&\times&
\qty(\frac{\mbh}{\num{20}\msun})^{0.28} \qty(\frac{\XI}{4})^{-0.28} \qty(\frac{\mcr}{20\msun})^{0.36} \label{eq:rgw} 
\earr
\esub
(see Appendix~\ref{appendix:rmin} for the derivation). This value is in a good agreement with \citet[][Figure 11]{Gondan2017}. 

However, previous papers \citep{OLeary2009,Kocsis2012} have assumed different definitions of $\rmin$, as summarized in Table~\ref{table:comparison}. 
\footnote{The reason for $\mSMBH$ dependence being different for different $\mtot$ in \citet{Gondan2017} (as well as in this work) is that for low mass BHs whose density declines less steeply than $r^{-2}$, most of the mergers are contributed by the largest $r$ rather than the smallest ones.}
In \cite{Kocsis2012} $\rmin$ is the radius with only one BH inside it (as determined by $n(r)$). However, even the region containing $<1$ BH {\it on average} can still make a non-negligible contribution to the total merger rate due to its very high average BH density and orbital velocity. We extrapolate the number density equation~\eqref{eq:n(m,r)} into this region given that $t_{\rm rel}\leq t_{\rm GW}$, but warn the reader that the assumptions used to derive that equation (phase space distribution function is smooth and correlations are negligible) break there.
And in \citet{OLeary2009} $\rmin$ is the radius where $t_{\rm GW}=t_{\rm H}$, which is a more conservative assumption than ours given that the relaxation time is smaller than $t_{\rm H}$. In any case, the dependence of the merger rate on $\rmin$ is rather weak ($\Gamma\propto\rmin^{-3/14}$).


\subsection{The effect of dynamical friction on the merger rate}

As the merger rate defined by Eq.~\eqref{eq:gamma0mamb} is proportional to the total numbers of BHs with masses $m_A$ and $m_B$, to account for the effects of DF we only have to multiply it by the corresponding BH number increase coefficients:

\begin{align}
 \frac{\partial^2\Gamma}{\partial m_A \partial m_B}  = \XI(m_A)\XI(m_B) \frac{\partial^2\Gamma_0}{\partial m_A \partial m_B} \label{eq:gammagamma0}
\end{align}

We present the two-dimensional (2D) mass distributions of the GW capture rate as a function of total BH mass and mass ratio following \citet{Gondan2017} and also calculate the marginalized 1D total mass distribution as discussed in Appendix \ref{appendix:ddGamma}.

\begin{figure*}
	\centering 
	\subfigure{\includegraphics[width=0.49\textwidth]{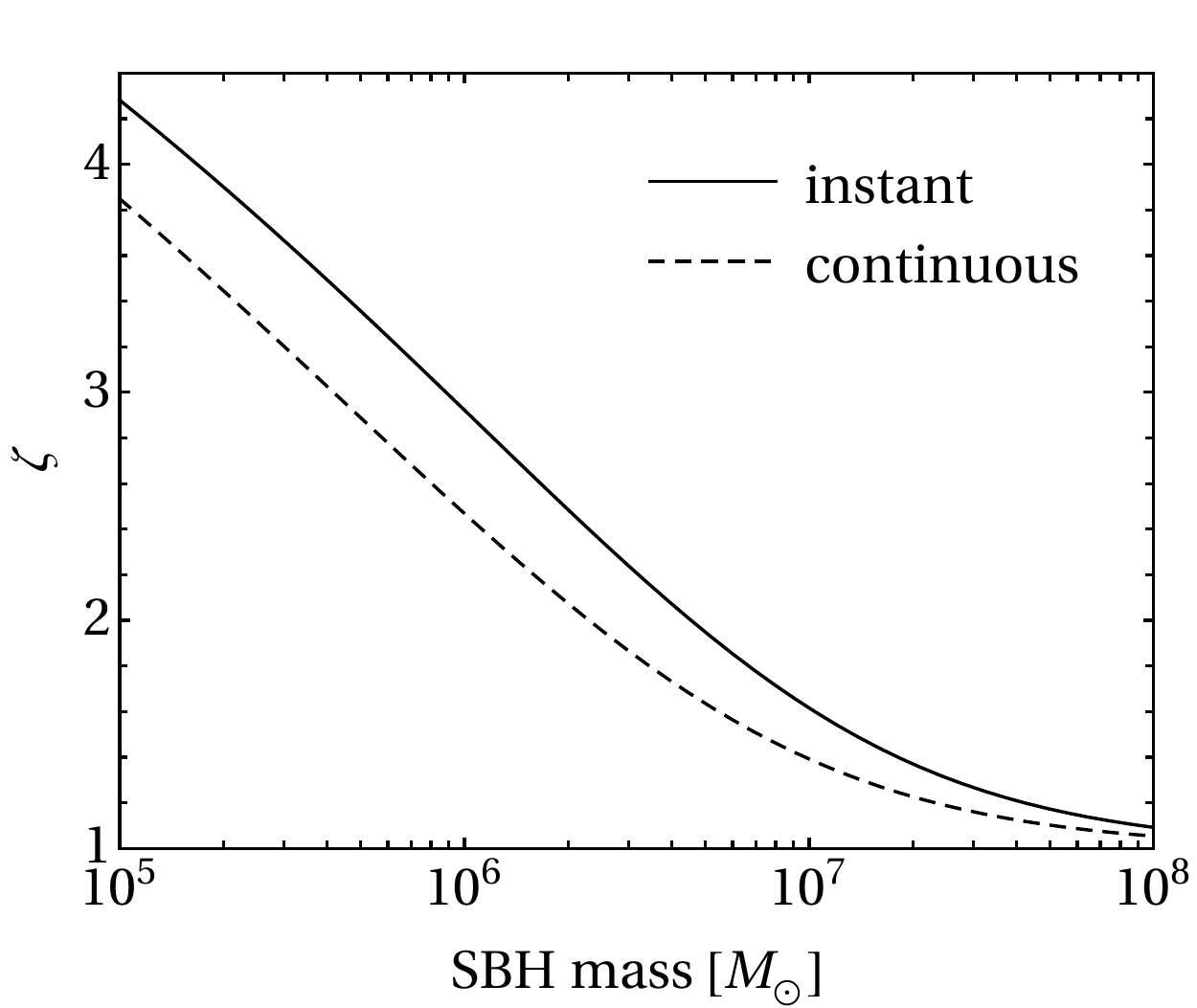}}
	\subfigure{\includegraphics[width=0.49\textwidth]{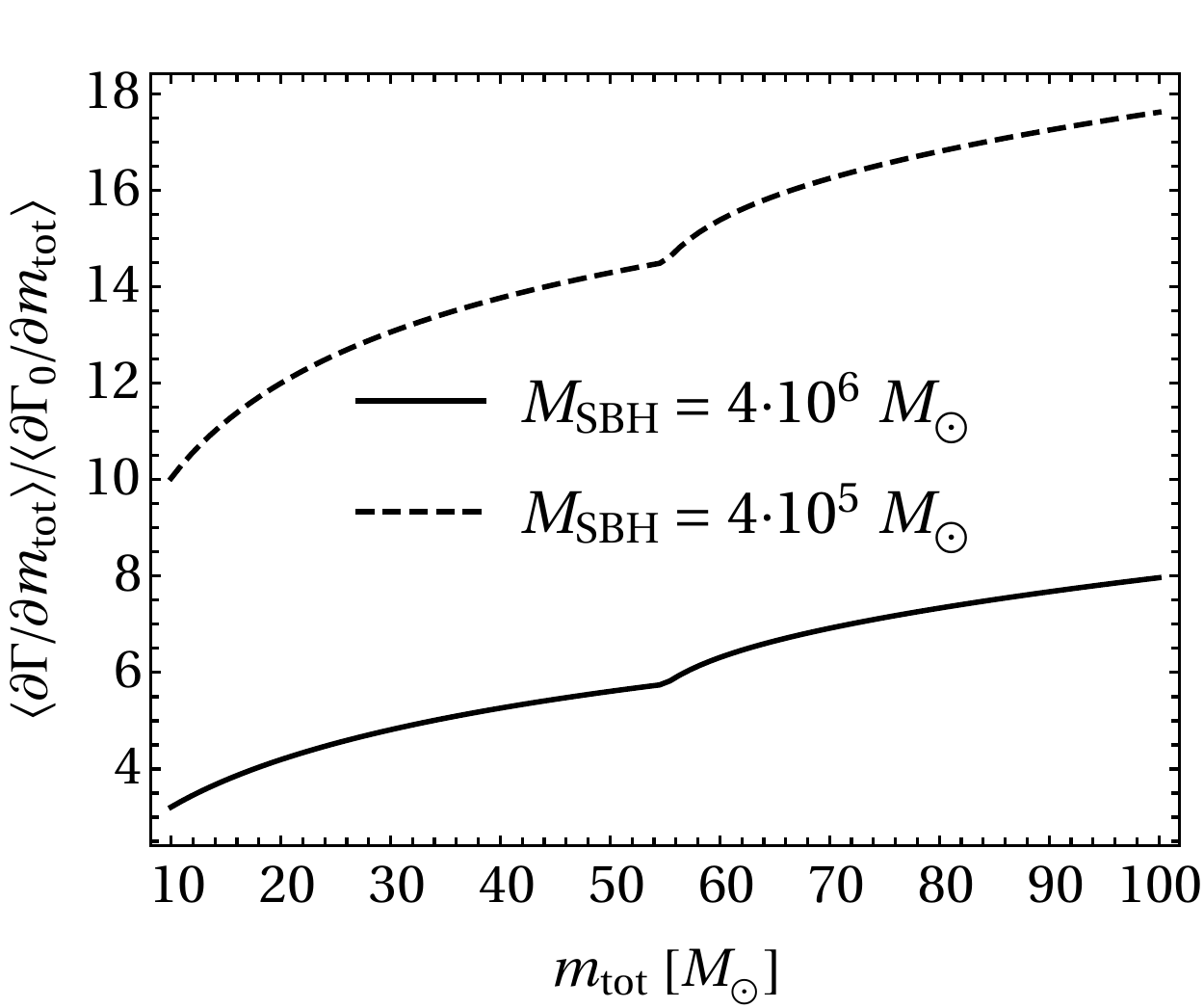}}
	\subfigure{\includegraphics[width=0.49\textwidth]{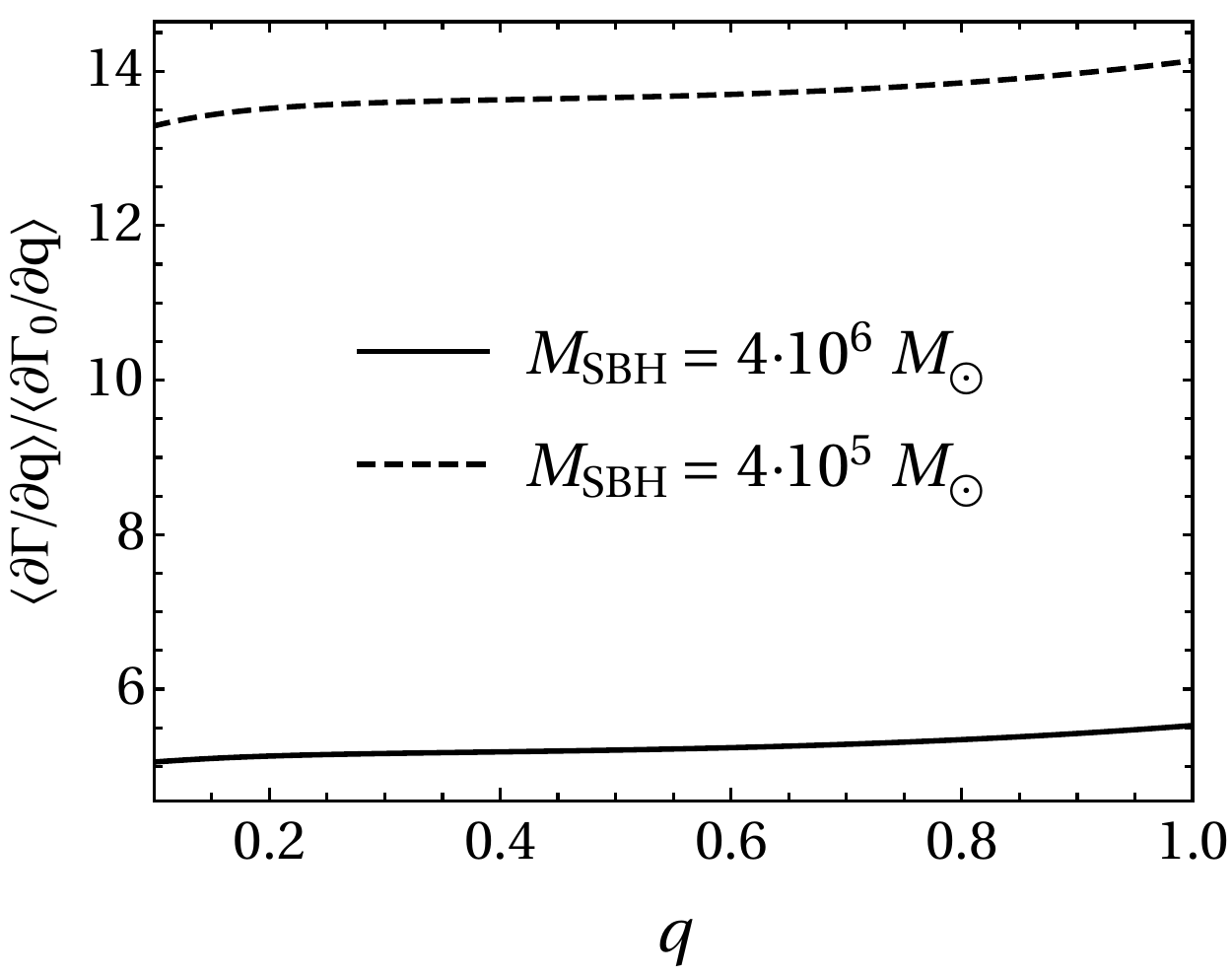}}
	\subfigure{\includegraphics[width=0.49\textwidth]{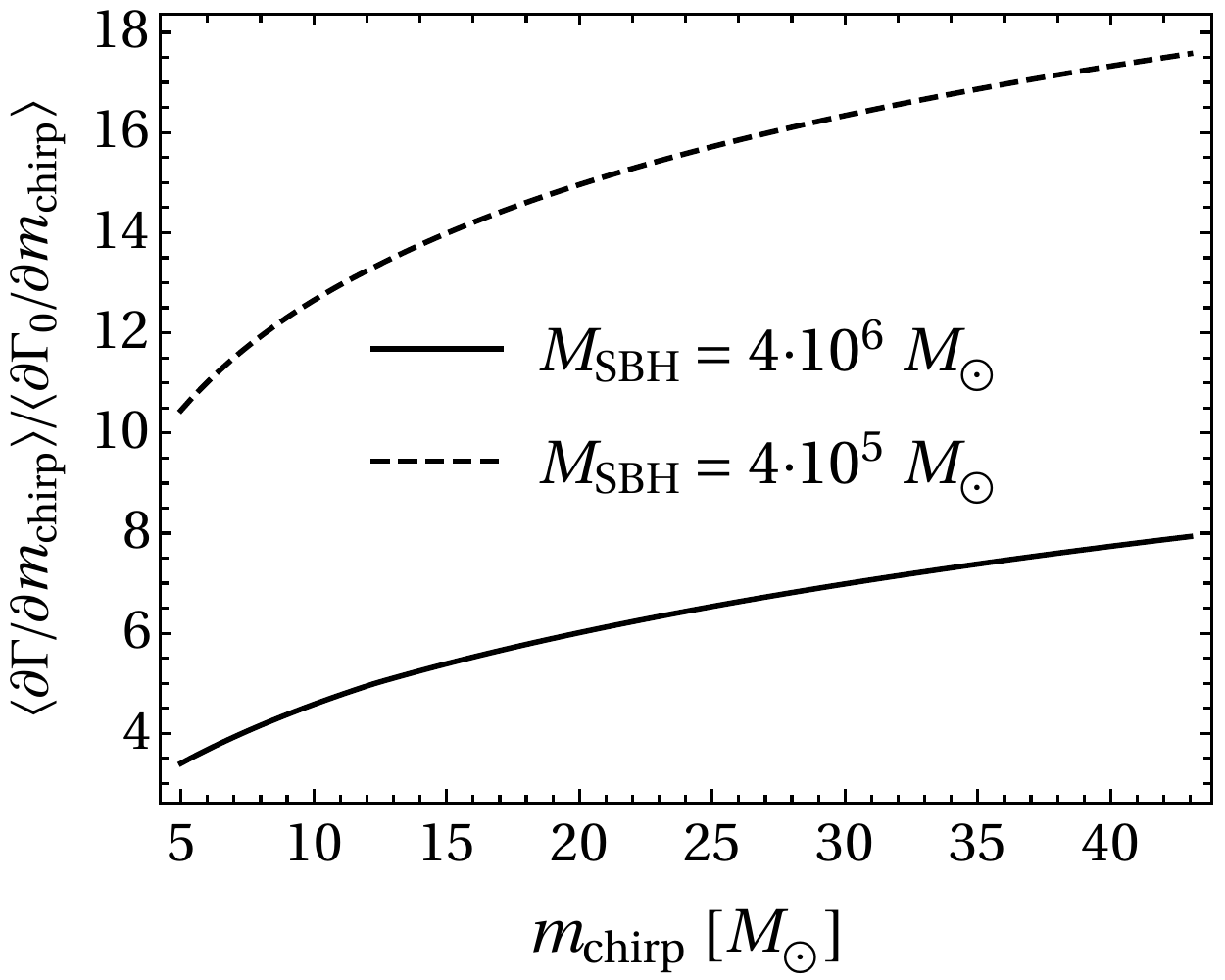}}
\caption{
Top left: the relative increase in BH population of the galactic center due to dynamical friction depending on the central SMBH mass and $m=10\msun$; all the other parameters are the same as in Figure~\ref{fig:xi}. Solid and dashed lines are for ``instantaneous'' and ``continuous'' BH formation models, respectively. The other panels show the relative increase in BH-BH merger rate for the instantaneous model of BH formation and two different values of the central SMBH mass  .
}
\label{fig:gg0}
\end{figure*}

To make a prediction for the total observed merger rate, we add up the local merger rates $\Gamma$ for every galaxy within the observable volume. For that purpose, it is useful to express $r_0$ in terms of the central supermassive BH mass using the $M-\sigma$ relation \citep{KormendyHo}:
\begin{align}
\label{eq:r_0}
&\mSMBH = M_0 \qty(\frac{\sigma}{\sigma_0})^{\alpha_0},\\
&M_0 = \num{3.097e8} \msun \qc \sigma=\SI{200}{km/s} \qc \alpha_0 = 4.384,\\
&r_0 = \frac{G\mSMBH}{\sigma^2} = \SI{3.14}{pc} \,\qty(\frac{\mSMBH}{\num{4e6}\msun})^{0.543}.
\end{align} 
The value of $r_0$ in this formula for a MW mass galaxy matches the measured value ($\approx\SI{3}{pc}$). 
{As for $\gamma_{1,2}$, }we assume they have MW values $\gamma_1=1.5$, $\gamma_2=3.2$ for all galaxies. Under these assumptions, Eq.~(\ref{eq:tdf0}) takes the form
\eq{\label{eq:tdf0-mSMBH}
t_\mathrm{DF,0} = \SI{4.1}{Gyr}\, \frac{1.5}{3-\gamma_1} \qty(\frac{m}{10\,\msun})^{-1} \qty(\frac{\mSMBH}{\num{4e6}\msun})^{1.31}
}
which is to be used in Eqs. \eqref{eq:rdf-outside}--\eqref{eq:rdf} instead of Eq.~\eqref{eq:tdf0}. 
This shows that the DF time in more massive galaxies is longer, implying that $\XI$, the increase in BH number due to DF, is less, as shown in Figure~\ref{fig:gg0} (top left).

In analogy with $\XI$, we calculate the relative increase in merger rate due to {the DF} effects, which is shown in Figure~\ref{fig:gg0} for two different values of $\mSMBH$. In accord with Eq.~(\ref{eq:tdf0-mSMBH}), we can see that the enhancement is stronger for heavier BHs and lower-mass galaxies.
As also shown in \citet{OLeary2009,Kocsis2012,Gondan2017}, the dependence of merger rate per galaxy on the SMBH mass (without explicitly taking into account DF effects) is very weak -- which implies that the total merger rate is dominated by numerous low-mass galaxies. And our results show that in these galaxies the merger rates per galaxy are up to 20 
times higher due to dynamical friction. This increases the overall contribution of low-mass galaxies even further.

\section{Total merger rate}
\label{section:total-merger-rate}

The total merger rate per unit volume may be calculated from the average merger rate per galaxy with a given SMBH mass and the distribution of the number density of galaxies with respect to the SMBH mass:
\barr
\frac{\partial^2\mathcal{R}}{\partial m_A \partial m_B} &=& \int_\mSMBHmin^\mSMBHmax \left\langle\frac{\partial^2\Gamma}{\partial m_A \partial m_B}\right\rangle 
\nonumber\\&\times&
\dv{n_\mathrm{gal}}{\mSMBH} \dd{\mSMBH}.
\earr
Previously we assumed that all galaxies follow the
$M-\sigma$ relation \eqref{eq:r_0}
exactly. 
In practice there may be significant variations in the model parameters between galaxies so that 
\begin{enumerate}[label=(\roman*)]
\item
\beq\label{i:mSMBH}
\mSMBH = C_{M\sigma} M_0 (\sigma/\sigma_0)^{\alpha_0},
\quad \langle C_{M\sigma}\rangle = 1,
\eeq
\item
the $r_0$ containing the stellar mass $M_\ast=2\mSMBH$ satisfies
\beq\label{i:r_0}
r_0 = \cinf \frac{G\mSMBH}{\sigma^2},
\eeq 
\item
\label{i:fBH}
the parameters of BH distribution ($m_\mathrm{min,max}$, $\beta$ and $\mcr$) as well as $\gamma_{1,2}$ may also vary over different galaxies, which also affect the total number of black holes in the NSC, $N_{\rm BH}$.
\end{enumerate}
Since $\left\{C_{M\sigma},\cinf,m_\mathrm{min},m_\mathrm{max}, \beta,\mcr,\gamma_1,\gamma_2\right\}$ may vary from galaxy to galaxy, this variance can significantly change (usually increase) the average merger rate compared to its value calculated using the average parameter values: 
\barr\label{eq:zeta}
&&\left\langle\frac{\partial^2\Gamma}{\partial m_A \partial m_B}(C_{M\sigma},\cinf,\dots)\right\rangle \nonumber\\&=& \xi_{M\sigma}\xi_\mathrm{inf} \xi_\mathrm{other} 
\frac{\partial^2\Gamma}{\partial m_A \partial m_B}\qty(\langle C_{M\sigma}\rangle,\left\langle\cinf\right\rangle,\dots),
\earr
where $\xi_{M\sigma}$, $\xi_\mathrm{inf}$ and $\xi_\mathrm{other}$ are the enhancement coefficients due to the variance in $C_{M\sigma}$, $C_\mathrm{inf}$ and all the other factors, respectively:
\barr
\xi_x \equiv \frac{\left\langle\frac{\partial^2\Gamma}{\partial m_A \partial m_B}(C_x)\right\rangle}{\frac{\partial^2\Gamma}{\partial m_A \partial m_B}\qty(\langle C_x\rangle)}.
\earr
Here we have assumed there are no correlations between different galaxy parameters and that the dependence of $\Gamma$ on them is separable:
\barr
\frac{\partial^2\Gamma}{\partial m_A \partial m_B} = f_1(C_{M\sigma}) f_2(\cinf) \dots
\earr


To obtain the parameter dependencies we eliminate $\sigma$ from the definition of $r_0$ in \eqref{i:r_0} using \eqref{i:mSMBH}, and substitute the result in Eqs.~\eqref{eq:rgw} and \eqref{eq:tdf0} to obtain the scaling of $r_{\min}$ and $\tdf$ with $C_{M\sigma}$ and $C_{\rm inf}$:
\bsub\label{eq:cmsigma}
\begin{align}
r_0 &= C_{\rm inf} G\mSMBH \qty(\frac{\mSMBH}{C_{M\sigma}M_0})^{-2/\alpha_0}\\
\rmin &\propto r_0^{0.31} \propto C_\mathrm{inf}^{0.31} C_{M\sigma}^{0.14},
\\
\tdfo &\propto r_0^{3/2} \propto C_{\rm inf}^{3/2} C_{M\sigma}^{0.68}.
\end{align}
\esub

According to \cite{KormendyHo}, the intrinsic scatter of $M-\sigma$ relation is 0.29 dex.
This implies $1.3\lesssim \xi_{M\sigma} \lesssim 1.5 $ depending on $m$ and $\mSMBH$ (Appexdix~\ref{appendix:msigma}).
As for $\cinf$, given the stellar density profile, it only depends on the velocity anisotropy {for relaxed NSCs} \citep[][Section 4.8.1]{BinneyTremaine}. We assume all galactic nuclei to be isotropic, which gives $\cinf=1$ and $\xi_{\rm inf}=1$. 


\citet{Tsang2013} made an estimate $\cinf=6.1$ based on the observed relation between 
$\rho_0$ and $\sigma$ and its scatter \citep{Merritt2007}. However, that is likely an upper limit to $\cinf$ as they have ignored the possible observational errors in both $\rho_0$ and $\sigma$ and also overestimated the spread in $\rho_0$ at fixed $\sigma$. 
Using the same plot from \citet{Merritt2007}, \citet{OLeary2009} arrived at the rough estimate of $\xi=30$; however, they only accounted for the variance in $\rho_0$ and ignored the variance in $r_0$ which is in fact related to $\rho_0$ at a given $\mSMBH$ (Eq. \ref{eq:rho0}).

The mass distribution of SMBHs is taken from \cite{Shankar2004}\footnote{It is consistent within uncertainties with the other SMBH mass estimates in the literature \citep[e.g.][]{Hopkins2007} as well as the SMBH masses inferred from the galaxy bulge mass distribution \citep{Thanjavur2016} with $\mSMBH/M_{\rm bulge}=0.003$.}:
\bsub\label{eq:dndM}
\barr
\dv{\ngal}{\mSMBH} &=& \frac{1}{\mSMBH\ln{10}} \dv{\ngal}{\log\mSMBH} \nonumber\\&=& 
\frac{\Phi_\ast}{\mSMBH\ln{10}} \qty(\frac{\mSMBH}{M_\ast})^{\alpha+1} \nonumber\\&\times&
\exp\qty[-\qty(\frac{\mSMBH}{M_\ast})^\beta],\\
\Phi_\ast &=& \SI{7.7e-3}{Mpc^{-3}}\qc\nonumber\\
M_\ast &=& \SI{6.4e7}{\msun}\qc\nonumber\\ \alpha&=&-1.11\qc\nonumber\\ \beta&=&0.49.
\earr
\esub
Instead of merger rate per unit $m_A$ and $m_B$, we calculate a more illustrative merger rate per unit log total mass and mass ratio:
\bsub
\begin{align}
&\frac{\partial^2\mathcal{R}}{\partial\log{\mtot} \partial q} = 
\frac{m_{\rm tot}^2}{(1+q)^2} \frac{\partial^2\mathcal{R}}{\partial m_A \partial m_B}\\ 
&= \SI{8.4e-16}{yr^{-1}}\, \frac{\xi}{1.4} \qty(\frac{\mtot/\msun}{1+q})^2 
\nonumber\\&\times
\qty(9-6p_0\frac{\mtot}{\mmax}+4p_0^2\frac{\mu\mtot}{m_\mathrm{max}^2})\\
&\times \frac{ m_\mathrm{tot}^{2-\beta} \mu^{-\beta} (1-\beta)^2 }{  \qty(m_\mathrm{max}
 ^{1-\beta} - m_\mathrm{min}^{1-\beta} )^2} 
\frac{c_\eta^{2/7}}{\frac{11}{14} - p_0 \frac{ m_\mathrm{tot} }{ m_\mathrm{max} }} 
\qty(\frac{\mcr}{20\msun})^{-2.6} \\
&\times \int_\mSMBHmin^\mSMBHmax 
\XI\qty(\frac{\mtot}{1+q},\mSMBH) \XI\qty(\frac{q\mtot}{1+q},\mSMBH)
\nonumber\\&\times
\qty(\frac{\mSMBH}{\num{4e6}\msun})^{3/28}\\
&\times \qty[1-\qty(\frac{\rmin}{r_0})^{\frac{11}{14} - p_0 \frac{ m_\mathrm{tot} }{ m_\mathrm{max} }}] \dv{n_\mathrm{gal}}{\mSMBH} \dd{\mSMBH},\\
\frac{\rmin}{r_0} &= \num{2.6e-5} \qty(\frac{\mSMBH}{\num{4e6}\msun})^{0.35} 
\qty(\frac{\mtot/(1+q)}{20\msun})^{0.28},\\
\xi &\equiv \xi_{M\sigma}\xi_\mathrm{inf} \xi_\mathrm{other}.
\end{align}
\esub

\begin{figure*}
	\centering
	\subfigure{\includegraphics[width=0.49\textwidth]{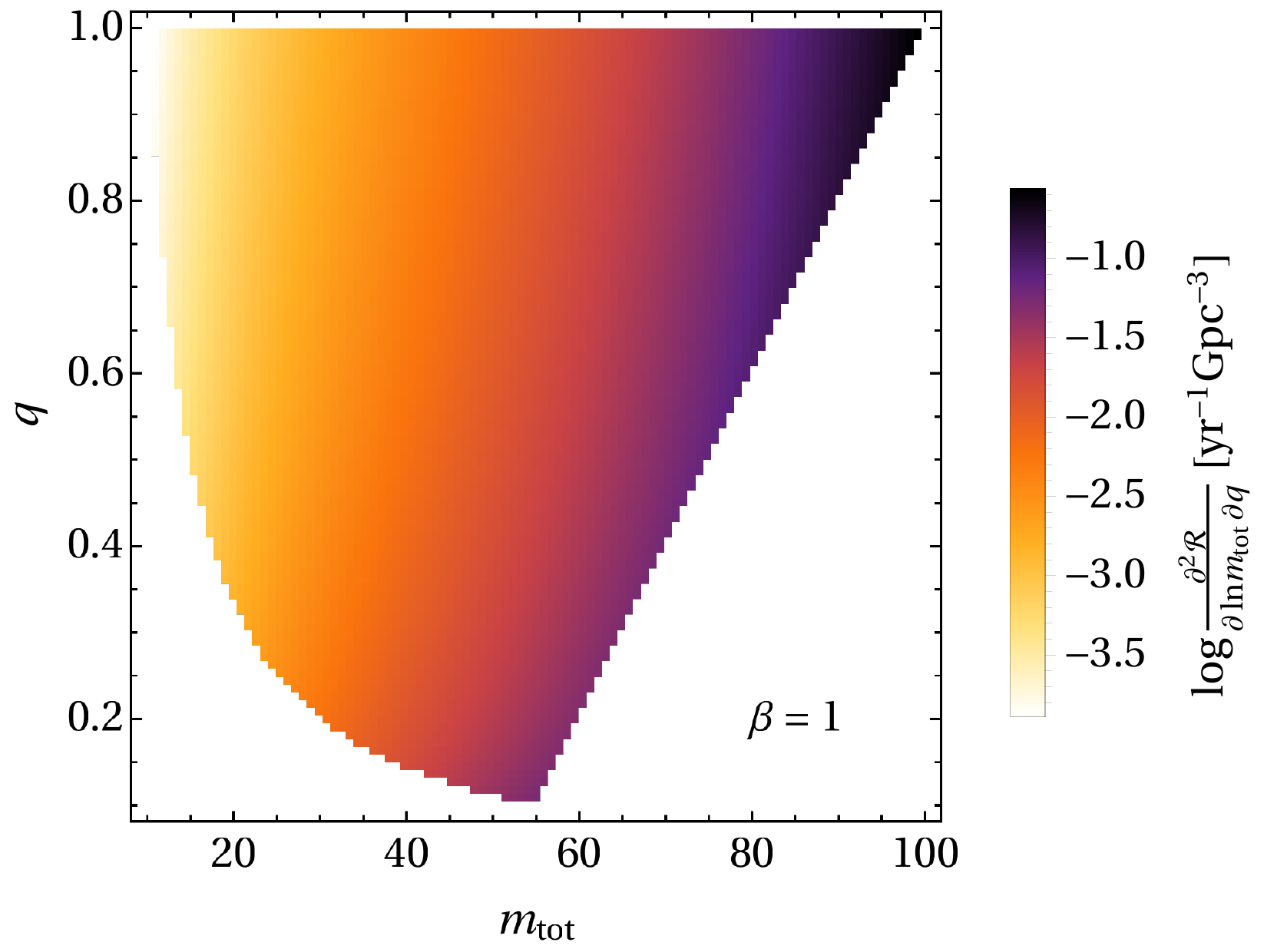}}
	\subfigure{\includegraphics[width=0.49\textwidth]{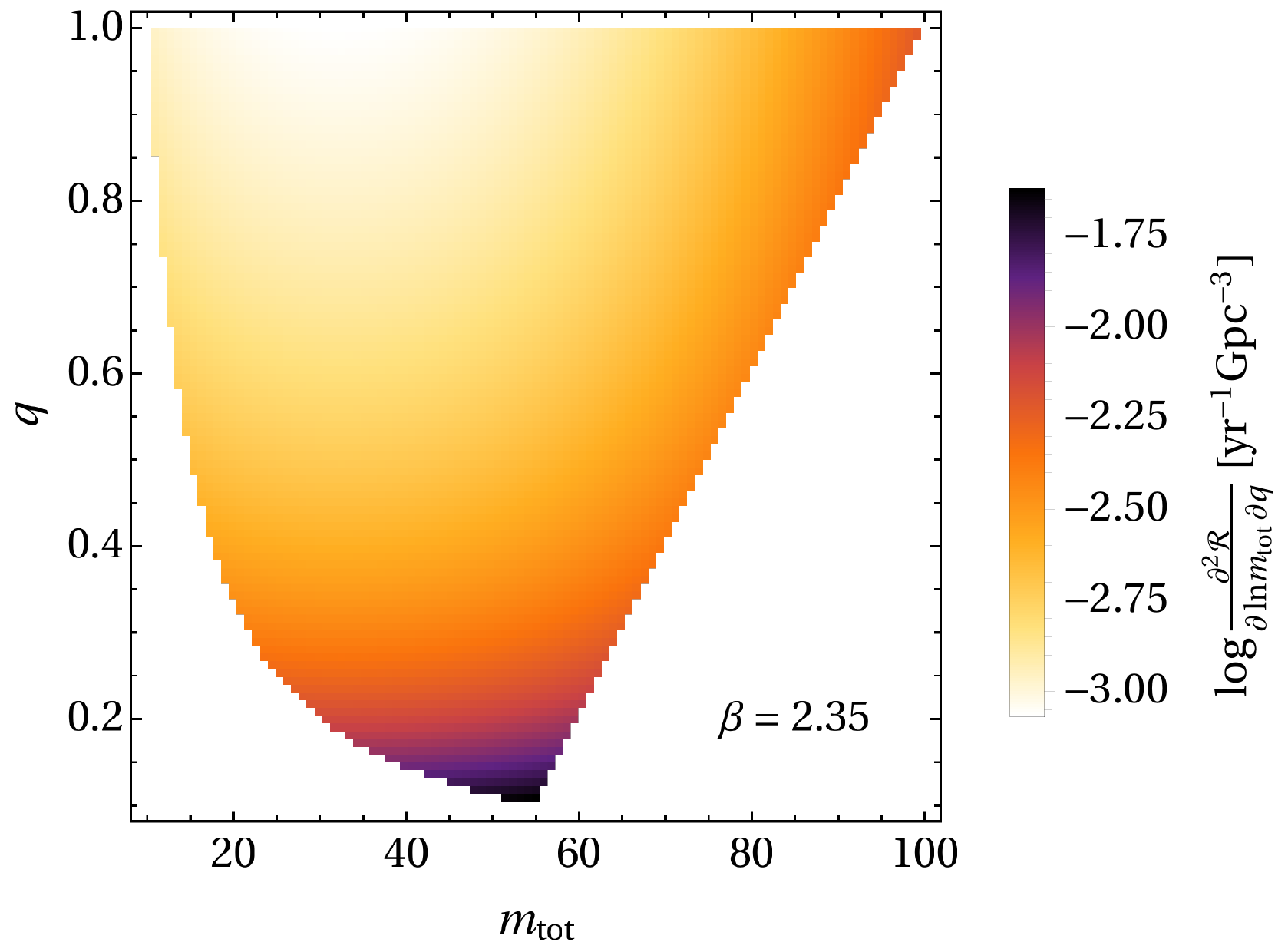}}
	\subfigure{\includegraphics[width=0.49\textwidth]{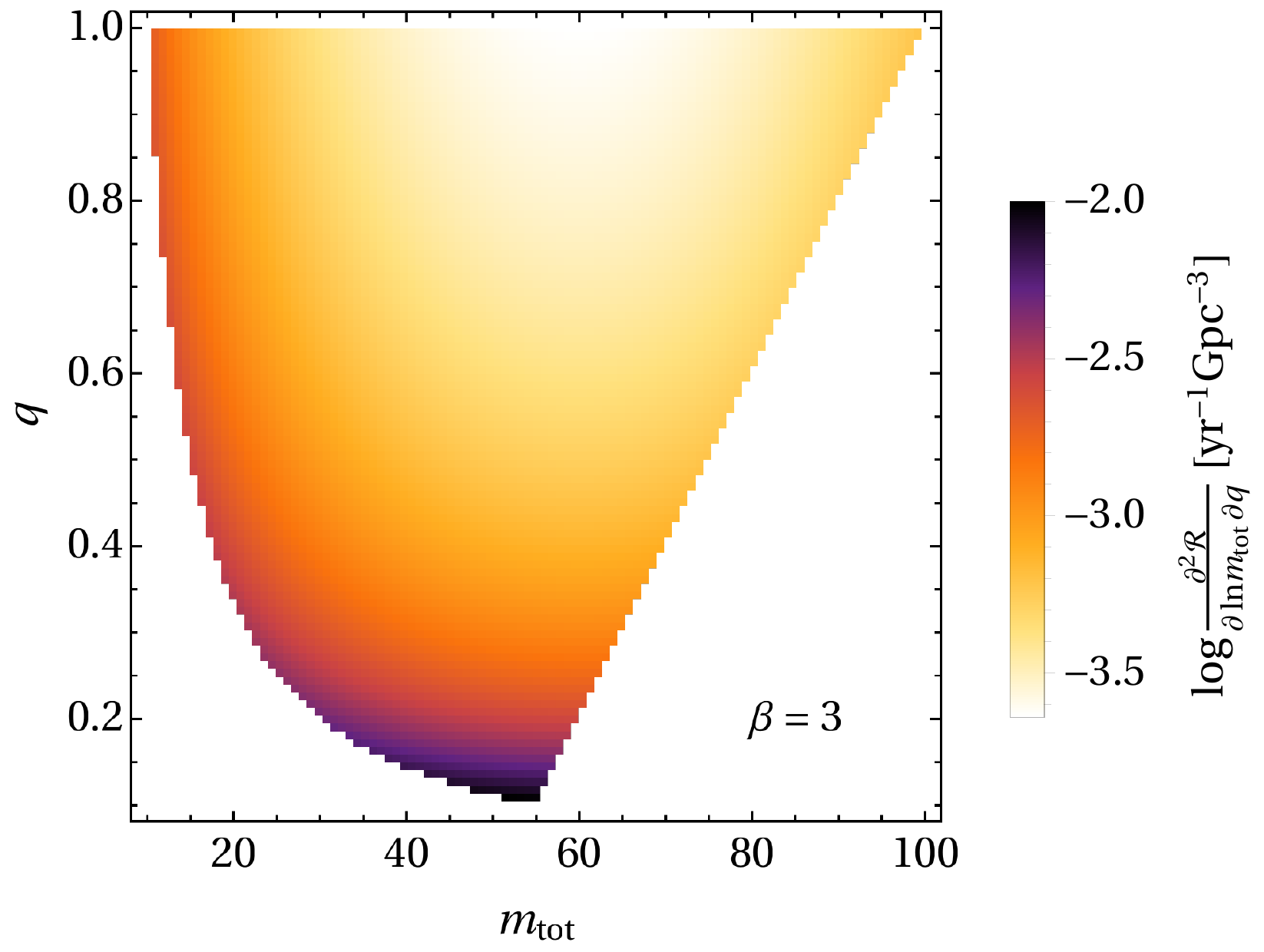}}
\caption{
Total merger rate per unit mass ratio per unit log total mass for BH mass distribution slope $\beta=1$, 2.35 and 3, respectively, assuming the minimal SMBH mass $\mSMBHmin=10^5\msun$.
}
\label{fig:mergerrate}
\end{figure*}

\begin{figure*}
	\centering
	\subfigure{\includegraphics[width=0.49\textwidth]{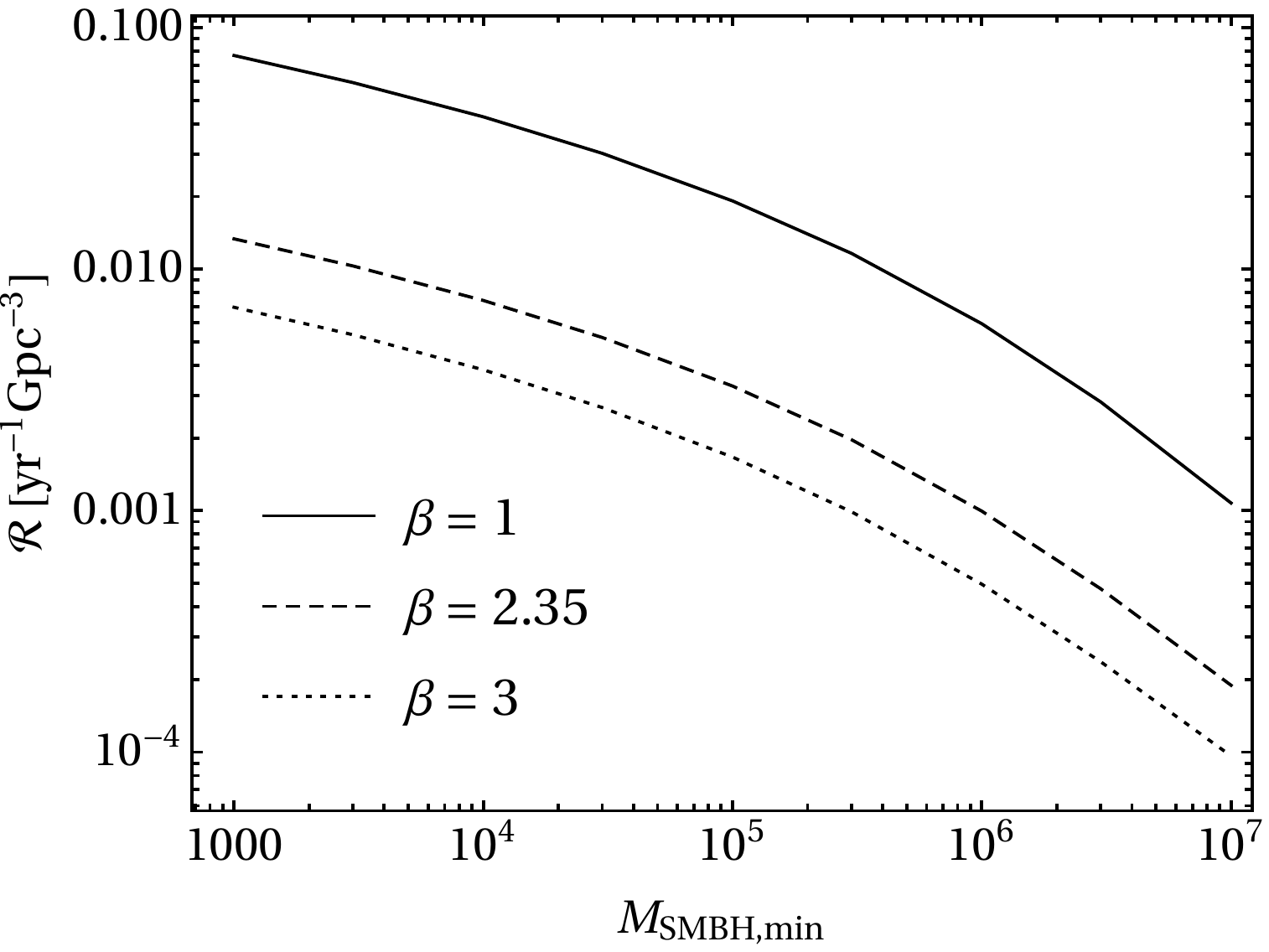}}
	\subfigure{\includegraphics[width=0.49\textwidth]{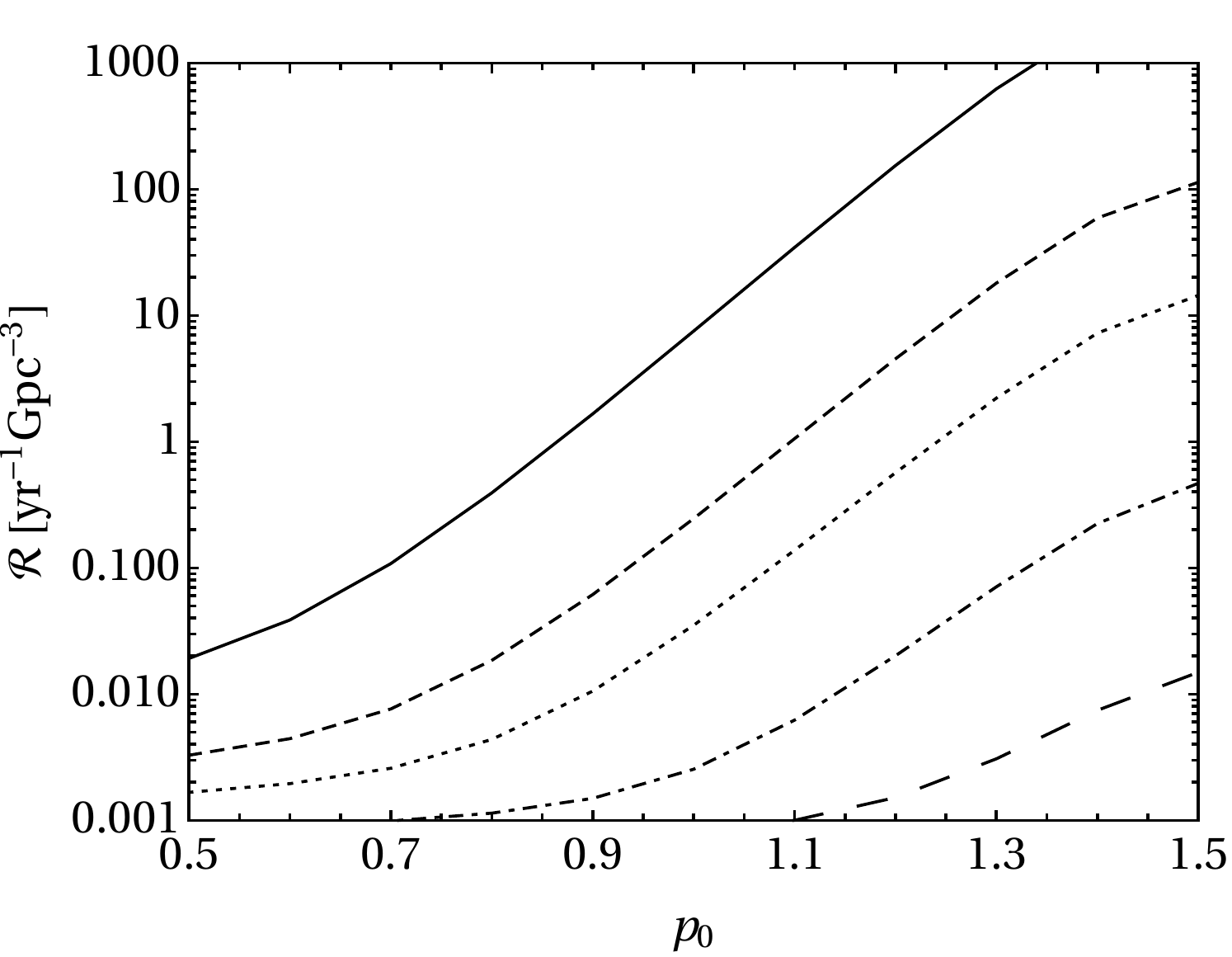}}
\caption{
Left: total merger rate (for all BH masses) depending on the assumption about the minimal SMBH mass in the Universe; different lines are for different values of $\beta$. Right: the dependence of total merger rate on the BH density profile (Eq.~\ref{eq:n(m,r)}) assuming $\xi_{M\sigma} = 1.4$, $\xi_{\rm inf} = \xi_{\rm other} = 1$, $M_\mathrm{SMBH,min} = 10^5\msun$. From top to bottom: $\beta=1,\,2.35,\,3,\,4,\,5$.
}
\label{fig:totalmergerrate}
\end{figure*}

The results of the calculation are shown in Figure~\ref{fig:mergerrate}. Compared to Figure~9 of \cite{Gondan2017}, the distribution is shifted towards higher BH masses due to DF. The SMBH mass function is almost log-uniform at low SMBH masses, and low-mass galaxies actually contribute more events due to enhanced DF (see Figure~\ref{fig:gg0}, top left). 
Therefore, the total merger rate depends on how far into the low SMBH masses does the distribution \eqref{eq:dndM} extend, which is illustrated in Figure~\ref{fig:totalmergerrate} (left). 
This figure shows the total merger rate integrated over all BH masses and mass ratios:
\eq{
\mathcal{R} = \int_{2\mmin}^{2\mmax}\dd{\mtot} \int_{\mmin/\mmax}^1\dd{q}
\frac{\partial^2\mathcal{R}}{\partial\log{\mtot} \partial q}
}
Figure \ref{fig:totalmergerrate} (left panel) also shows that the merger rate is a decreasing function of $\beta$ (i.e. an increasing function of the average BH mass).
Given $\xi=3$, $p_0=0.5$, $\mcr=20\msun$, and ranges $M_\mathrm{SMBH,min}=10^4-10^5\msun$ and $\beta=1-3$, the overall merger rate is 
$0.002-\SI{0.04}{Gpc^{-3}yr^{-1}}$.

Based on fits to the isotropic Fokker Planck results of \citet{OLeary2009}, which imply $p_0\approx0.5-0.6$, we have assumed $p_0=0.5$, i.e. a power-law BH radial density distribution given by Eq.~\eqref{eq:n(m,r)} which has a slope 1.5 for the lightest BHs and 2 for the heaviest ones. However, using different mass functions in isotropic Fokker-Planck models, \citet[][Figure 3]{Keshet2009} have shown that for $\beta\gtrsim4$ and $\mmax/\mmin\gtrsim10$ the heaviest BH density profile can be steeper, up to $r^{-3}$, which would correspond to $p_0=1.5$ in Eqs. \eqref{eq:n(m,r)} and \eqref{eq:gamma0mamb}. 
There is indeed some observational evidence that the surface density distribution of massive O-stars (i.e. BH progenitors) in the Galactic center is $\propto R^{-1.4}$ \citep{Bartko2009}, which implies 3D density $\propto r^{-2.4}$, i.e. $p_0\sim0.9$ \citep[see, however,][who claim that the young star density distribution is better described by broken power-law with the inner slope $\propto R^{-0.9}$]{Stostad2015}.
Another possible reason for the density cusp to be steeper than $r^{-2}$ is binary disruption by the SMBH's tidal field \citep{FragioneSari2018}. Figure~\ref{fig:totalmergerrate} (right) shows that such an increase in $p_0$ could increase the merger rate by orders of magnitude and therefore deserves further study.

Up to this point, we have assumed the BH density distribution to be spherical. However, the MW NSC is observed to be flattened with mean axis ratio 0.7-0.8 \citep{Schodel2014,Fritz2016}. \citet{Feldmeier+17} found it to be decreasing towards the center within $r<\SI{1}{pc}$, reaching 0.4 when $r\rightarrow0$. According to their orbit-based modelling, that corresponds to a triaxial density profile with axial ratios $c/a=0.28$, $b/a=0.64$ in the center. The triaxiality leads to increase in the average BH density $n_{\rm BH}$ within $r_0$ and, consequently, to an increase in the event rate $\mathcal{R}\propto n_{\rm BH}^2$. A crude upper limit estimate for this increase in $\mathcal{R}$ may be obtained as $\qty(a^2/bc)^2\approx30$. In addition, from a theoretical point of view, vector resonant relaxation in a multimass system causes the heaviest objects (BHs) to segregate from an initially spherical stellar distribution into a disk \citep{Szolgyen+18}. The final distribution of heavy BH angular momenta in Fig. 2 of \citet{Szolgyen+18} implies the $\mathcal{R}$ increase by a factor of $\approx2.8$. How effective that phenomenon is in non-spherical systems is currently unknown and deserves further study.

\begin{figure}
	\centering
	\subfigure{\includegraphics[width=0.49\textwidth]{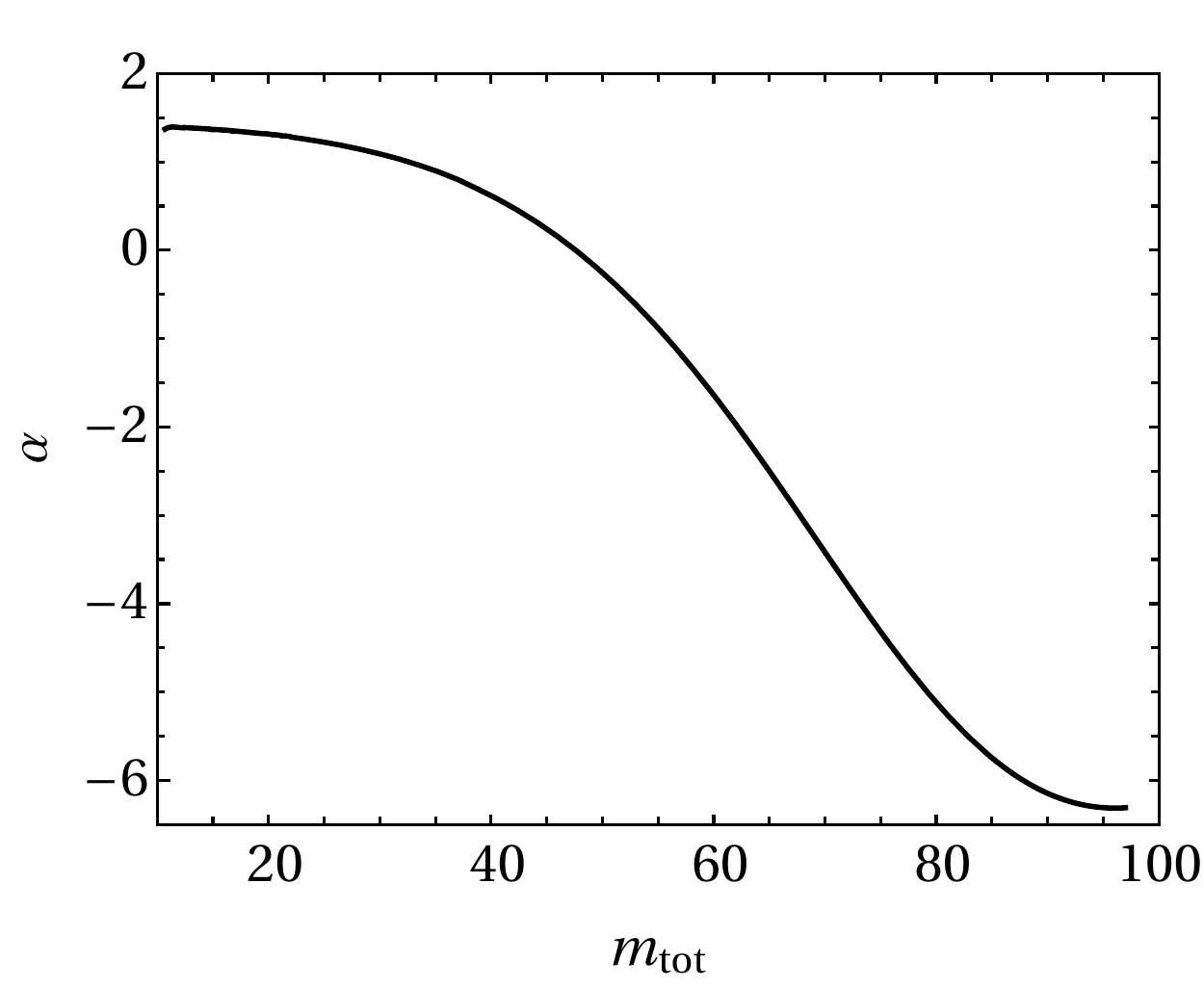}}
	\subfigure{\includegraphics[width=0.49\textwidth]{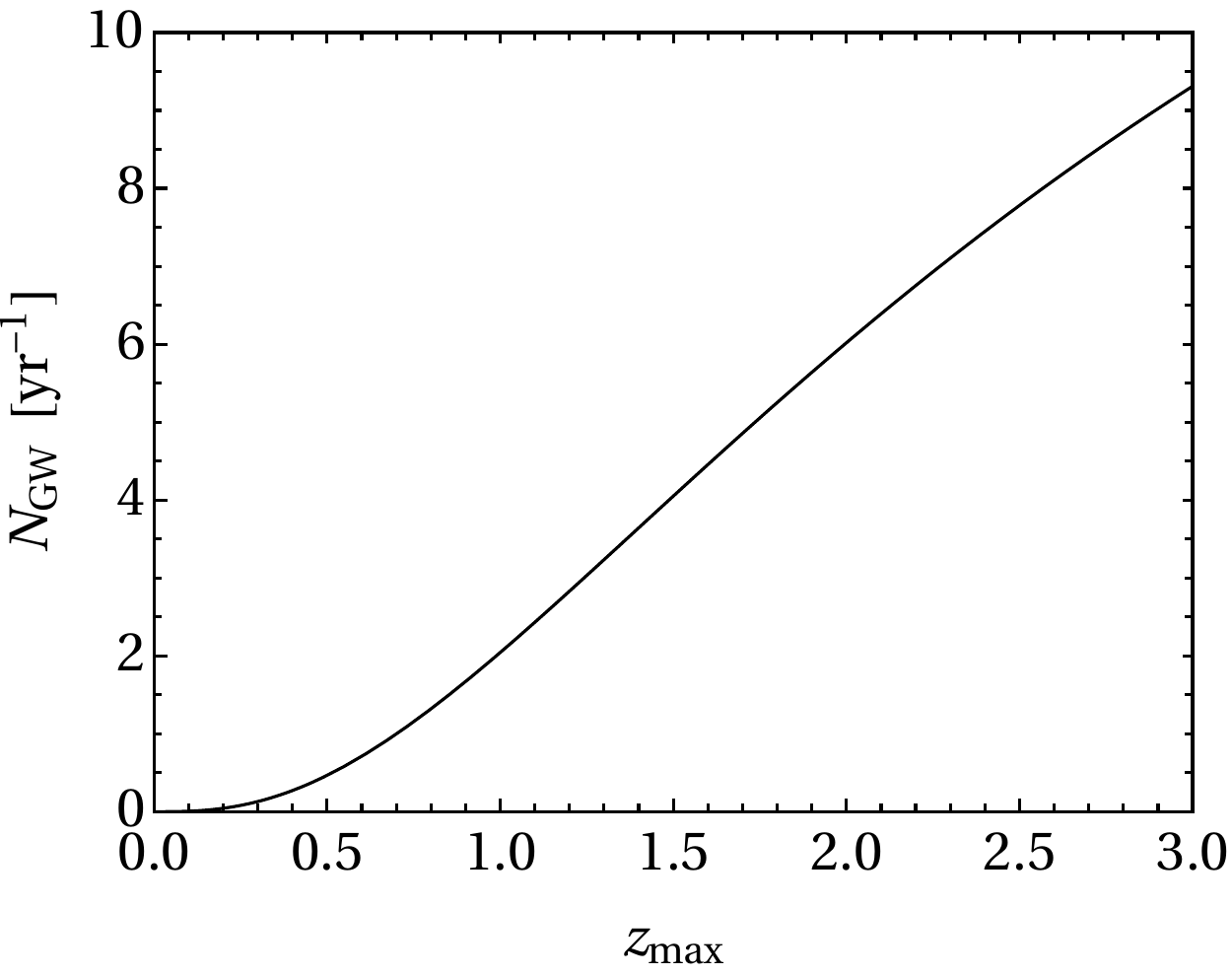}}
\caption{
Top: dimensionless parameter $\alpha$ characterizing the BH population (Eq. \ref{eq:alpha}) at various values of the total mass of the binary. Bottom: the expected detection rate assuming maximum detection redshift $z_\mathrm{max}$ and merger rate $\SI{0.02}{Gpc^{-3}yr^{-1}}$. 
}
\label{fig:alpha}
\end{figure}

In addition to the merger rate, we also calculate the universal dimensionless parameter
\barr\label{eq:alpha}
\alpha = -\mtot^2 \pdv{}{m_A}{m_B} \ln\pdv{\mathcal{R}}{m_A}{m_B}
\earr
which is independent of the BH mass function and is sensitive to the astrophysical process leading to the BH merger \citep{Kocsis2018}. Its value varies from 1.4 for the smallest BHs to $-6.3$ for the heaviest ones which is in good agreement with \citet{Gondan2017}; its dependence on $\mtot$ is shown in Figure~\ref{fig:alpha} (top). It turns out to be practically independent of $q$. This is different from, e.g., $\alpha=1$ for primordial BH binaries formed in the early universe \citep{Kocsis2018} or $\alpha=1.43$ for BHs in dark matter halos \citep{Bird2016}.

\section{Conclusions}
\label{section:conclusions}

We have calculated the number density of stellar-mass BHs in a galactic nucleus with a SMBH in the center using simplified isotropic models, where we assumed the BHs and stars reach a mass-segregated steady-state distribution inside the influence radius and took into account the dynamical friction bringing BHs inside that region. We have shown that dynamical friction can increase the BH number up to $\sim5$ times and also that its effect is much more pronounced in small galaxies.

We used this information to calculate the rate of GW captures in galactic nuclei taking into account the observed SMBH mass distribution and the scaling relations between the influence radius, SMBH mass and velocity dispersion. The total event rate turns out to be dominated by small galaxies, due to both the event rate per galaxy being weakly dependent on mass and the dynamical friction effect being more pronounced in smaller galactic nuclei.

The event rate is determined, on one hand, by the BH mass distribution parameters (their total mass fraction $\kappa$ and mass distribution slope $\beta$) and also by the SMBH mass function below $\sim10^7\msun$: the SMBH number density per unit $\log\mSMBH$ per unit comoving volume $N_\mathrm{SMBH}$ and the SMBH mass lower limit $M_\mathrm{SMBH, min}$. The approximate dependence of the total event rate on all these parameters is 
\barr\label{eq:final}
\mathcal{R} &\approx& \SI{0.02}{Gpc^{-3}yr^{-1}} 
\frac{N_\mathrm{SMBH}}{\SI{0.015}{Gpc^{-3}dex^{-1}}} \qty(\frac{\kappa}{\num{2.9e-3}})^2 
\nonumber\\&\times&
\qty(\frac{\mSMBHmin}{10^5\,\msun})^{-0.32} e^{1.06(1-\beta)},
\earr
which is 
similar to a previously published estimate 
of \citet{Tsang2013} and below the rates cited in \citet{OLeary2009}, $0.6\,(\xi/30)\dots 45\,(\xi/30)\,\si{yr^{-1}Gpc^{-3}}$ for different galaxy models. However, Eq. \eqref{eq:final} assumes a certain BH number density distribution (Eq. \ref{eq:n(m,r)} with $p_0=1/2$); a steeper ($p_0>1/2$) and/or non-spherical distribution could increase $\mathcal{R}$ by orders of magnitude, as discussed in Section \ref{section:total-merger-rate}.


The GW capture rates we calculated are much lower than the current estimates by LIGO \citep[$24-112\,\si{yr^{-1}Gpc^{-3}}$,][]{ligo2018}, therefore they're unlikely to be the dominant source of BH mergers. The number of predicted aLIGO detections per year in our model is determined by the event rate per unit comoving volume $\mathcal{R}$ and the accessible volume:
\bsub
\barr
\ngw &=& \mathcal{R} \int_0^{z_\mathrm{max}} \frac{1}{1+z} \dv{V_c}{z} \dd{z},\\
\dv{V_c}{z} &=& 4\pi \frac{c}{H_0}\frac{d^2(z)}{\sqrt{\Omega_M(1+z)^3+\Omega_\Lambda}},\\
d(z) &=& \frac{c}{H_0}\int_0^z\frac{\dd{z'}}{\sqrt{\Omega_M(1+z')^3+\Omega_\Lambda}}.
\earr
\esub
Here $z_\mathrm{max}$ is the maximum accessible redshift (we ignore its dependence on the BH masses and other parameters of the binary), $\dv*{V_c}{z}$ is the comoving volume per unit redshift and $d(z)$ is the comoving distance. 
Figure~\ref{fig:alpha} (bottom) illustrates that dependence. 
In this equation we have not accounted for the possibility of $p_0$ in Eq.~(\ref{eq:n(m,r)}) being higher than 0.5 that could potentially increase the event rate up to a few orders of magnitude (Figure~\ref{fig:totalmergerrate}, right).

A possible way to distinguish GW captures from the other channels is their high eccentricity in the LIGO frequency range ($e>0.1$ at $f>\SI{10}{Hz}$).  However, eccentric mergers can also be produced in triple systems where a BH binary achieves extreme eccentricity through interaction with with a third body; those systems could be
\begin{itemize}
\item A binary BH orbiting around a SMBH in a galactic center can experience variations of inclination and eccentricity due to Kozai-Lidov effect, sometimes reaching $e>0.9999$ which quickly leads to coalescence through GW energy loss \citep{Antonini2012,FragioneKL,Hamers2018,Hoang2018}. 
\item Lidov-Kozai effect can also cause the inner BH binary coalescence in hierarchical BH triples which form through binary-binary interactions in globular clusters
\citep{Antonini2016} or through the evolution of massive star triples in galactic field \citep{Antonini2017,Silsbee2017}.
\item Triple BHs can also form via binary BH -- single BH encounters in globular clusters \citep{Samsing+2018,Zevin2019}. A merger subsequently happens as a result of GW emission during a close encounter between two of the three BHs while they temporarily form a bound three-body state \citep{Samsing2018}.
\end{itemize}
Table \ref{table:mergerRates} shows the merger rates for all those different mechanisms in different environments. The rates turn out to be comparable ($\sim0.01-0.1\,\si{Gpc^{-3}yr^{-1}}$) but it still might be possible to distinguish them by their eccentricity distribution.

\begin{table*}
\caption{Merger rate density of events with $e>0.1$ in the LIGO band ($>\SI{10}{Hz}$), $\si{Gpc^{-3}yr^{-1}}$}
\centering
\renewcommand{\arraystretch}{2}
\renewcommand\cellgape{\Gape[4pt]}
\begin{tabular}{| c | c | c | c |}
\hline
& GW capture (single-single interactions) & Hierarchical truples (Kozai-Lidov effect) & Binary-single interactions \\
\hline
Nuclear star clusters & \makecell{$0.002-0.04$\footnote{Assuming $p_0=0.5$; up to $\SI{1.7}{Gpc^{-3}yr^{-1}}$ assuming a stronger mass segregation with $p_0=0.9$.} (this work) \\ $0.6-45$ \citep{OLeary2009} \\ 0.02 \citep{Tsang2013} } & ? & ? \\
\hline
Globular clusters & ? & 0.04 \citep{Antonini2016} & 0.5 \citep{Rodriguez2018} \\
\hline
Galactic field & 0? & \makecell{$0.002-0.1$ \citep{Silsbee2017} \\ $0.01-0.04$ \citep{Antonini2017}} & 0?\\
\hline
\end{tabular}
\label{table:mergerRates}
\end{table*}

Apart from NSCs with a SMBH in the center we considered in this paper, other kinds of star clusters could also contribute to the GW capture merger rate: NSCs without SMBHs, globular clusters and open clusters. However, in those other systems the velocity dispersion is not as high which means the GW capture mergers in them are much less eccentric \citep[Figure 6 in][]{OLeary2009}. It's also worth noticing that the number density of NSCs with SMBHs in them can be up to $40\%$ higher due to ultracompact dwarf galaxies some of which  are believed to be stripped NSCs of low-mass galaxies \citep{Voggel2018}. Finally, we assumed all BHs to be formed in-situ an haven't accounted for the BHs brough into the Galctic Center via globular cluster inspiral \citep{ArcaSedda1,ArcaSedda2}.

\section*{Acknowledgements}

This work received funding from the European Research Council (ERC) under the European Union's Horizon 2020 research and innovation programme under grant agreement No 638435 (GalNUC) and was supported by the Hungarian National Research, Development, and Innovation Office grant NKFIH KH-125675.

\appendix
\section{Increase in BH number in the assumption of continuous BH formation}
\label{appendix:xi-continuous}

Let $t(m,r)$ be the time it takes for a BH of mass $m$ to reach $r_0$ starting from $r>r_0$. From Eq.~(\ref{eq:rdf-outside}) we know that 
\eq{
t = \tdfo \qty(\frac{(r/r_0)^{1/k_3}-1}{k_1})^{1/k_2},
}
where $k_{1,2,3}$ are the same as in Eq.~(\ref{eq:rdf-outside}).
Assuming BHs are formed with a constant rate, only a fraction $1-t/T$ of all the BHs formed at radius $r$ will be born early enough to reach $r_0$ by present time. Therefore, Eq.~(\ref{eq:ntot}) for the total number of BHs at present now reads as follows:
\barr
\dv{\ntot}{m} &=& \fbhi \int_0^{r_0} n(r)\,4\pi r^2\dd{r} + \fbhi \int_{r_0}^{\rdf(m)} n(r)\,4\pi r^2\dd{r} \qty(1-\frac{t(m,r)}{t_H}),
\earr
and the BH number increase coefficient is 
\eq{
\XI &= 1 + (3-\gamma_1) \int_1^{\rdf/r_0} x^{2-\gamma_2} \qty(1-\frac{\tdfo}{T}\qty(\frac{x^{1/k_3}-1}{k_1})^{1/k_2}) \dd{x}.
}
\section{Minimum radius of BH distribution}
\label{appendix:rmin}

As mentioned in Section~\ref{section:mergerrate}, the BHs are depleted due to their inspiral into the SMBH below a minimum radius $\rgw$:
\barr\label{eq:rlx=gw}
0.34\frac{\sigma^3(r_\mathrm{GW})}{G^2n(r_\mathrm{GW})\langle m^2\rangle\ln\Lambda} = \frac{5c^5\rgw^4}{64G^3\mbh\mSMBH^2}
\earr
where {$\sigma(r)=\sqrt{G\mSMBH/r}$ and} $n$ is the total number density of objects (both BHs and stars) and $\langle m^2\rangle$ is their average squared mass:
\barr\label{eq:nm2}
n(r)\langle m^2\rangle = \int_0^\infty \dv{n_\ast(r,m_\ast)}{m_\ast} m_\ast^2 \dd{m_\ast} + \int_\mmin^\mmax \dv{n_\mathrm{BH}(r,\mbh)}{\mbh} \mbh^2 \dd{\mbh} = n_\ast(r)\langle m_\ast^2\rangle + n_\mathrm{BH}(r) \langle \mbh^2\rangle
\earr
At $r\ll r_0$, which is a reasonable assumption given Eq.~(\ref{eq:rlx=gw}), the BH term dominates the right-hand side of Eq.~(\ref{eq:nm2}). Indeed,
\barr\label{eq:dn/dm}
\dv{n_\mathrm{BH}}{m} = \qty(\frac{r}{r_0})^{-3/2-p_0m/\mmax} C(m)\,,
\earr
where $C(m)$ is a function of $m$ that can be found from
\barr
\int_0^{r_0}\dv{n_\mathrm{BH}}{m} 4\pi r^2\dd{r} = \dv{\nbh(r_0)}{m} = \XI(m) \dv{\ninit(r_0)}{m}\,,
\earr
$\ninit(r_0)$ and $\nbh(r_0)$ being the initial and final numbers of BHs within the sphere of influence (Eq.~\ref{eq:ninit}), so that
\barr\label{eq:dninit/dm}
\dv{\ninit(r_0)}{m} = \ninit(r_0) \frac{(1-\beta)\,m^{-\beta}}{\mmax^{1-\beta}-\mmin^{1-\beta}}\,.
\earr


Eqs. (\ref{eq:dn/dm})--(\ref{eq:dninit/dm}) yield
\barr
\dv{n_\mathrm{BH}}{m} = \frac{\XI\ninit(r_0)}{4\pi r_0^3} \qty(\frac{3}{2} - p_0\frac{m}{\mmax}) \frac{(1-\beta)\,m^{-\beta}}{\mmax^{1-\beta}-\mmin^{1-\beta}} \qty(\frac{r}{r_0})^{-3/2-p_0m/\mmax}
\earr
which implies that
\barr
n_\mathrm{BH}(r) \langle \mbh^2\rangle \approx \frac{\XI\ninit(r_0)}{4\pi r_0^3} \frac{1-\beta}{\mmax^{1-\beta}-\mmin^{1-\beta}} \int_\mmin^\mmax \qty(\frac{3}{2} - p_0\frac{m}{\mmax}) \qty(\frac{r}{r_0})^{-3/2-p_0m/\mmax} m^{2-\beta} \dd{m}\,.
\earr
{Here we have ignored the dependence of $\XI$ on BH mass. This is justified by the limited range of $\XI$ (Fig.~\ref{fig:xi}), weak dependence of $\rmin$ on $\XI$ (Eq.~\ref{eq:rgw}) and also weak dependence of the merger rate on $\rmin$ (Eq.~\ref{eq:gamma0mamb}). For example, for a SBH mass $10^6\,\msun$ the assumption of constant $\XI=4$ results in $<10\%$ error in the value of $n_\mathrm{BH}(r) \langle \mbh^2\rangle$.}
At the default values of $p_0=0.5$, $\mmax=40\msun$, $\mmin=5\msun$ and $\beta=2.3$ the dependence of the integral on $r/r_0$ can be approximated by a power law {(with $3\%$ accuracy for $r/r_0<10^{-3}$ which turns out to be a safe assumption)}:
\barr\label{eq:nbh}
n_\mathrm{BH}(r) \langle \mbh^2\rangle \approx \frac{\XI\ninit(r_0)}{4\pi r_0^3} \cdot 116\msun^2 \qty(\frac{r}{r_0})^{-1.88}.
\earr
	

For stellar number density we have
\barr
n_\ast(r)\langle m_*^2\rangle &=& \frac{\rho_\ast \langle m_*^2\rangle}{\langle m_\ast\rangle} = \frac{3\mSMBH \langle m_*^2\rangle}{4\pi r_0^3\langle m_\ast\rangle }\qty(\frac{r}{r_0})^{-3/2}
\earr
From Eq.~(\ref{eq:KroupaIMF}) $\langle m_\ast\rangle = 0.30\msun$ and $\langle m_\ast^2\rangle = 0.79\msun^2$ 
, so that 
\barr
\frac{n_\mathrm{BH}(r) \langle \mbh^2\rangle}{n_\ast(r)\langle m_\ast^2\rangle} = \qty(\frac{r}{r_0})^{-0.38}\frac{\XI\ninit(r_0)\langle m_\ast\rangle}{3\mSMBH}\frac{116\msun^2}{\langle m_\ast^2\rangle} \approx 
0.33\frac{\XI}{4} \qty(\frac{r}{r_0})^{-0.38} \qty(\frac{\mcr}{20\msun})^{-1.3}.
\earr
We can conclude that BHs dominate the relaxation process everywhere inside $0.05r_0$ 
\cite[similarly to what was reported in][Fig. 2]{OLeary2009}, and the stellar term can indeed be ignored in Eq.~(\ref{eq:rlx=gw}). Combined with Eqs. (\ref{eq:nbh}) and (\ref{eq:lnLambda}), Eq. (\ref{eq:rlx=gw}) yields Eq.~(\ref{eq:rgw}). 


\section{Contribution of BHs inside $\rmin$ to the event rate}
\label{appendix:insidermin}

Inside the sphere of radius $\rmin$ where relaxation is ineffective compared to the GW-induced orbital radius shrinking (the orbits are assumed circular) the BH number density is defined by
\barr
\dv{N}{r} = 4\pi r^2 n(r)\propto \dv{t}{r} \propto r^5, 
\earr 
so that $n(r) \propto r^3$. From \cite{Kocsis2012} we know that the merger rate per unit radius
\barr
\dv{\Gamma}{r} \propto r^{39/14} n^2(r)
\earr
Here we consider mergers between heaviest BHs that have $n(r)\propto r^{-2}$, $r>\rmin$. Then
\barr
\dv{\Gamma}{r} = C
\begin{dcases}
\qty(\frac{r}{\rgw})^{-17/14}\qc r>\rgw\\
\qty(\frac{r}{\rgw})^{123/14}\qc r<\rgw
\end{dcases}\qc
C = \mathrm{const}.
\earr
From this we can conclude that the contribution of BHs inside $\rmin$ to the total event rate is negligible:
\barr
\frac{\Gamma(r<\rmin)}{\Gamma(r>\rmin)} = 
\frac{\int_0^{\rmin}\dv{\Gamma}{r}\dd{r}}{\int_{\rmin}^{\infty}\dv{\Gamma}{r}\dd{r}} = 
\frac{(14/137)C\rgw}{(14/3)C\rgw} \approx 0.02
\earr

\section{The impact of the intrinsic scatter of $M-\sigma$ relation}
\label{appendix:msigma}

The increase in the merger rate $\Gamma$ due to scatter in $C_{M\sigma}$ (distributed log-normally with mean 1 and standard deviation $\delta$) is
\barr\label{eq:zetamsigma}
\xi_{M\sigma}=
\qty[\int_0^{\infty} \frac{\partial^2\Gamma(C_{M\sigma})}{\partial m_A \partial m_B}  
\exp(-\frac{(\ln{C_{M\sigma}})^2}{2\delta^2}) \frac{\dd{C_{M\sigma}}}{\sqrt{2\pi}\delta C_{M\sigma}}]\bigg/
\qty[\frac{\partial^2\Gamma}{\partial m_A \partial m_B} (1)]
\earr
As we can see from Eqs. \eqref{eq:gamma0mamb} and \eqref{eq:gammagamma0}, $\Gamma$ depends on $C_{M\sigma}$ through $r_0$, $\rmin$ and $\XI$:
\bsub
\barr
\frac{\partial^2\Gamma(C_{M\sigma})}{\partial m_A \partial m_B} &\propto& \XI(m_A)\,\XI(m_B)\,r_0^{-31/14} \qty|1 - \qty(\frac{r_\mathrm{min}}{r_0})^{11/14 - p_0 m_\mathrm{tot} / m_\mathrm{max}}| \nonumber\\
&\propto& \XI(m_A)\,\XI(m_B)\times
\begin{dcases}
C_{M\sigma}^{-1.26+0.16\mtot/\mmax}\qc m_\mathrm{tot} > \frac{11}{7} m_\mathrm{max}\\
C_{M\sigma}^{-1.01}\qc m_\mathrm{tot} < \frac{11}{7} m_\mathrm{max}
\end{dcases},\\
\XI &=& 1 + \frac{3-\gamma_1}{\gamma_2-3}
\qty[1-\qty{1+k_1\qty(2.9\frac{m}{10\,\msun} \qty(\frac{\mSMBH}{\num{4e6}\msun})^{-1.31}C_{M\sigma}^{-0.68})^{k_2}}^{(3-\gamma_2)k_3}].
\earr
\esub
Here we used Eqs. (\ref{eq:cmsigma}), (\ref{eq:rdf}) and (\ref{eq:tdf0-mSMBH}) to determine the dependence of $r_0$, $\rmin$ and $\XI$ on $C_{M\sigma}$.
Calculating the integral in Eq.~(\ref{eq:zetamsigma}) numerically, we see that $\xi_{M\sigma}$ is a decreasing function of $m_{A,B}$ and at $\delta=0.29\ln10$ it spans {a range of values $\xi_{M\sigma}=1.3-1.5$ for $m_{A,B}\in[5\msun,50\msun]$, $\mSMBH\in[10^5\msun,10^7\msun]$.} 

\section{Manipulation of merger rate distributions}
\label{appendix:ddGamma}

As derived in \cite{Gondan2017}, the merger rate distributions per unit $\mtot$, $q\equiv m_B/m_A$, $\mathcal{M}$ and $\eta$ can be calculated as

\bsub
\begin{align}
 \frac{\partial^2\Gamma}{\partial \mtot \partial q}  &= \frac{\mtot}{(1+q)^2} \frac{\partial^2\Gamma}{\partial m_A \partial m_B} , \\
 \frac{\partial^2\Gamma}{\partial \mathcal{M} \partial \eta}  &= \mathcal{M}\eta^{-6/5}(1-4\eta)^{-1/2} \frac{\partial^2\Gamma}{\partial m_A \partial m_B} .
\end{align}
\esub

The merger rate distribution as a function of only one variable can be given by marginalizing one of these equations over the other variable, e.g.
\begin{align}
     \frac{\partial\Gamma}{\partial \mtot}  = \int^1_\frac{\mmin}{\mmax}  \frac{\partial^2\Gamma}{\partial \mtot q}  \dd{q} 
 = \int^1_\frac{\mmin}{\mmax}  \frac{\partial^2\Gamma_0}{\partial m_A \partial m_B}  \frac{\mtot}{(1+q)^2} 
 \XI\qty(\frac{\mtot}{1+q}) \XI\qty(\frac{q\mtot}{1+q}) \dd{q},
\end{align}
where $\Gamma_0$ is the merger rate calculated without taking into account the effects of DF (Eqs.~\ref{eq:gamma0mamb}).

\bibliographystyle{yahapj}
\bibliography{bib}

\end{document}